\documentclass[pdftex,twocolumn,epjc3]{svjour3}          
\usepackage[english]{babel}

\makeatletter
\let\cl@chapter\undefined
\makeatletter

\RequirePackage[T1]{fontenc}
\smartqed  
\RequirePackage{graphicx}
\usepackage{newtxtext,newtxmath}
\RequirePackage{flushend}
\RequirePackage[numbers,sort&compress]{natbib}
\RequirePackage[colorlinks,citecolor=blue,urlcolor=blue,linkcolor=blue]{hyperref}

\journalname{Eur.\ Phys.\ J.\ C}

\usepackage[switch]{lineno}
\usepackage{diagbox}
\usepackage{tabularx}
\usepackage[figuresright]{rotating}
\usepackage{textcomp}
\usepackage{rotating,booktabs}
\usepackage[capitalise]{cleveref}
\usepackage{siunitx}
\usepackage{textpos}

\graphicspath{{figs/}}

\begin{document}

\title{The energy spectrum of cosmic rays beyond the turn-down around $\boldsymbol{10^{17}}$\,eV as measured with the surface detector of the Pierre Auger Observatory}

\titlerunning{Cosmic ray energy spectrum above $10^{17}$\,eV}        

\author{The Pierre Auger Collaboration\thanksref{email,addr}}

\thankstext{email}{e-mail: spokespersons@auger.org}

\institute{The Pierre Auger Observatory, Av.\ San Mart\'{\i}n Norte 306, 5613 Malarg\"ue, Mendoza, Argentina; http://www.auger.org\label{addr}}

\date{April 2022}

\maketitle

\begin{textblock}{15}(0,-1.35)
\noindent Published in Eur. Phys. J. C as DOI: \href{https://www.doi.org/10.1140/epjc/s10052-021-09700-w}{10.1140/epjc/s10052-021-09700-w}\\
\noindent Report Number: FERMILAB-PUB-21-474-AD-AE-SCD-TD
\end{textblock}

\begin{abstract}
We present a measurement of the cosmic-ray spectrum above 100\,PeV using the part of the surface detector of the Pierre Auger Observatory that has a spacing of 750~m. An inflection of the spectrum is observed, confirming the presence of the so-called \emph{second-knee} feature. The spectrum is then combined with that of the 1500\,m array to produce a single measurement of the flux, linking this spectral feature with the three additional breaks at the highest energies. The combined spectrum, with an energy scale set calorimetrically via fluorescence telescopes and using a single detector type, results in the most statistically and systematically precise measurement of spectral breaks yet obtained. These measurements are critical for furthering our understanding of the highest energy cosmic rays.
\end{abstract}


\section{Introduction}

The steepening of the energy spectrum of cosmic rays (CRs) at around $10^{15.5}$\,eV, first reported in~\cite{Khristiansen}, is referred to as the ``knee'' feature. 
A widespread view for the origin of this bending is that it corresponds to the energy beyond which the efficiency of the accelerators of the bulk of Galactic CRs is steadily exhausted. 
The contribution of light elements to the all-particle spectrum, largely dominant at GeV energies, remains important up to the knee energy after which the heavier elements gradually take over up to a few $10^{17}$\,eV~\cite{Arqueros:1999uq,Fowler:2000si,Aglietta:2004np,Aglietta:2003hq,Garyaka:2007pf}. 
This fits with the long-standing model that the outer shock boundaries of expanding supernova remnants are the Galactic CR accelerators, see e.g.~\cite{Blasi:2013rva} for a review. Hydrogen is indeed the most abundant element in the interstellar medium that the shock waves sweep out, and particles are accelerated by diffusing in the moving magnetic heterogeneities in shocks accordingly to their rigidity. That the CR composition gets heavier for two decades in energy above the knee energy could thus reflect that heavier elements, although sub-dominant below the knee, are accelerated to higher energies, until the iron component falls off steeply at a point of turn-down around ${\simeq}10^{16.9}$\,eV. Such a bending has been observed in several experiments at a similar energy, referred to as the
``second knee'' or ``iron knee''~\cite{Apel:2012tda,IceCube:2019hmk,Abbasi:2018xsn,Budnev:2020oad}.
The recent observations of gamma rays of a few $10^{14}~$eV from decaying neutral pions, both from a direction coincident with a giant molecular cloud~\cite{Albert:2020yty} and from the Galactic plane~\cite{Amenomori:2021gmk}, provide evidence for CRs indeed accelerated to energies of several $10^{15}~$eV, and above, in the Galaxy.
A dozen of sources emitting gamma rays up to $10^{15}~$eV have even been reported~\cite{LhaasoNature}, and the production could be of hadronic origin in at least one of them~\cite{Cao:2021hdt}.
However, the nature of the sources and the mechanisms by which they accelerate CRs remain in general undecided.
In particular, that particles can be effectively accelerated to the rigidity of the second knee in supernova remnants is still under debate, see e.g.~\cite{Cristofari:2020mdf}.

Above $10^{17}$\,eV, the spectrum steepens in the interval leading up to the ``ankle'' energy, ${\sim}5{\times}10^{18}$\,eV, at which point it hardens once again. 
The inflection in this energy range is not as sharp as suggested by the energy limits reached in the Galactic sources to accelerate iron nuclei beyond the iron-knee energy~\cite{Hillas:2005cs}. 
Questions arise, then, on how to make up the all-particle spectrum until the ankle energy. 
The hardening around $10^{17.3}$\,eV in the light-particle spectrum reported in~\cite{Apel:2013uni} is suggestive of an extragalactic contribution to the all-particle spectrum steadily increasing.
It has even been argued that an additional component is necessary to account for the extended gradual fall-off of the spectrum and for the mass composition in the iron-knee-to-ankle region, be it of Galactic~\cite{Hillas:2005cs} or extragalactic origin~\cite{Aloisio:2013hya}.

While the concept that the Galactic-to-extragalactic transition occurs somewhere between $10^{17}$\,eV and a few $10^{18}$\,eV is well-accredited, a full understanding of how it occurs is hence lacking. 
The approximately power-law shape of the spectrum in this energy range may mask a complex superposition of different components and phenomena, the disentanglement of which rests on the measurements of the all-particle energy spectrum, and of the abundances of the different elements as a function of energy, both of them challenging from an experimental point of view. 
On the one hand, the energy range of interest is accessible only through indirect measurements of CRs via the extensive air showers that they produce in the atmosphere. 
Therefore, the determination of the properties of the CRs, especially their mass and energy, is prone to systematic effects. 
On the other hand, different experiments, different instruments and different techniques of analysis are used to cover this energy range, so that a unique view of the CRs is only possible by combining measurements the matching of which inevitably implies additional systematic effects. 

The aim of this paper is to present a  measurement of the CR spectrum from $10^{17}$\,eV up to the highest observed energies, based on the data collected  with the surface-detector array of the Pierre Auger Observatory. 
The Observatory is located in the Mendoza Province of Argentina at an altitude of 1400\,m above sea level at a latitude of $35.2^\circ$ S, so that the mean atmospheric overburden is 875\,g/cm$^2$. 
Extensive air showers induced by CR-interactions in the atmosphere are observed via a hybrid detection using a fluorescence detector (FD) and a surface detector (SD). 

\begin{figure}[t]
\centering
\includegraphics[width=0.99\columnwidth]{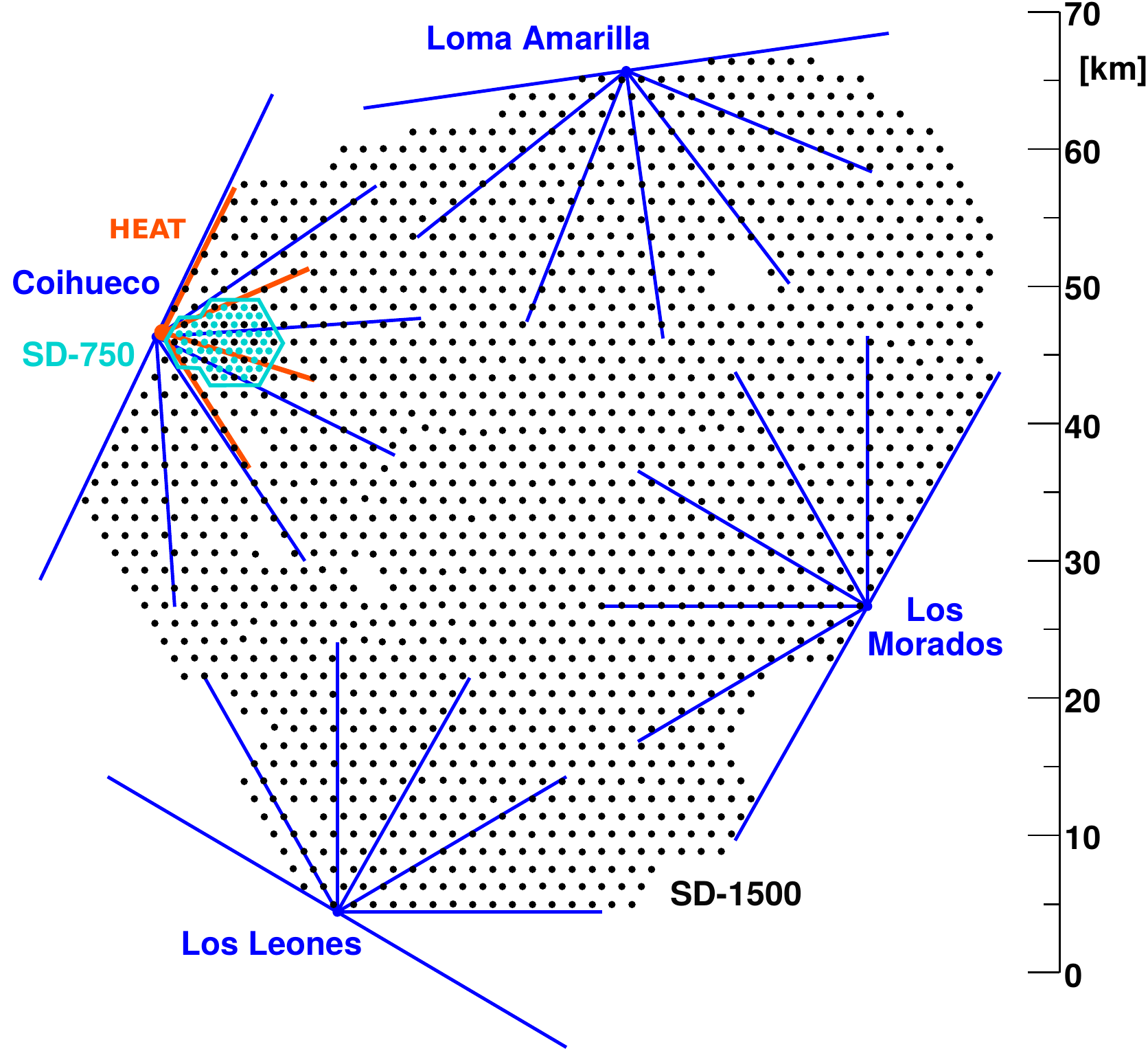}
\caption{The layout of the SD and FD of the Pierre Auger Observatory are shown above. The respective fields of view of the five FD sites are shown in blue and orange. The 1600 SD locations which make up the SD-1500 are shown in black while the stations which belong only to the SD-750 and the boarder of this sub-array are highlighted in cyan.}
\label{f:ArrayMap}
\end{figure}
The FD consists of five telescopes at four sites which look out over the surface array, see~\cref{f:ArrayMap}.
Four of the telescopes (shown in blue) cover an elevation range from $0^\circ$ to $30^\circ$ while the fifth, the High Elevation Auger Telescopes (HEAT), covers an elevation range from $30^\circ$ to $58^\circ$ (shown in red).
Each telescope is used to collect the light emitted from air molecules excited by charged particles. 
After first selecting the UV band with appropriate filters (310 to 390\,nm), the light is reflected off a spherical mirror onto a camera of 22$\times$20 hexagonal, 45.6\,mm, photo-multiplier tubes (PMTs). 
In this way, the longitudinal development of the particle cascades can be studied
and the energy contained within the electromagnetic sub-showers can be measured in a calorimetric way.
Thus the FD can be used to set an energy scale for the Observatory that is calorimetric and so is independent of simulations of shower development.

The SD, the data of which are the focus of this paper, consists of two nested hexagonal arrays of water Cherenkov detectors (WCDs). The layout, shown in~\cref{f:ArrayMap}, includes the SD-1500, with detectors spread apart by 1500\,m and totaling approximately 3000\,km$^2$ of effective area. 
The detectors of the SD-750 are instead spread out by 750\,m, yielding an effective area of 24\,km$^2$. SD-750 and SD-1500 include identical WCDs, cylindrical tanks of pure water with a 10\,m$^2$ base and a height of 1.2\,m. Three 9" PMTs are mounted to the top of each tank and view the water volume. When relativistic secondaries enter the water,
Cherenkov radiation is emitted, reflected via a Tyvek lining into the PMTs, and digitized using 40 MHz 10-bit Flash Analog to Digital Converters (FADCs).
Each WCD along with its digitizing electronics, communication hardware, GPS, etc., is referred to as a \emph{station}.

Using data collected over 15 years with the SD-1500, we recently reported the measurement of the CR energy spectrum in the range covering the region of the ankle up to the highest energies~\cite{Aab:2020rhr,Aab:2020gxe}. 
In this paper we extend these measurements down to $10^{17}$\,eV using data from the SD-750: not only is the detection technique consistent but the same methods are used to treat the data and build he spectrum. 
The paper is organized as follows: we first explain how, with the SD-750 array, the surface array is sensitive to primaries down to $10^{17}$\,eV in \cref{s:newtriggers}; in \cref{s:EnergyMeasurements}, we describe how we reconstruct the showers up to determining the energy; we illustrate in \cref{s:Spectrum} the approach used to derive the energy spectrum from SD-750; finally, after combining the spectra measured by SD-750 and SD-1500, we present the spectrum measured using the Auger Observatory from $10^{17}$\,eV upwards in \cref{s:combination} and discuss it in the context of other measurements in \cref{s:conclusion}.
\section{Identification of Showers with the SD-750: From the Trigger to the Data Set}
\label{s:newtriggers}

The implementation of an additional set of station-level trigger algorithms in mid-2013 is particularly relevant for the operation of the SD-750. Their inclusion in this work extends the energy range over which the SD-750 triggers with $>98\%$ probability from $10^{17.2}$\,eV down to $10^{17}$\,eV.

To identify showers, a hierarchical set of triggers is used which range in scope from the individual station-level up to the selection of events and the rejection of random coincidences. 
The trigger chain, extensively described in~\cite{Abraham:2010zz}, has been used since the start of the data taking of the SD-1500, and was successively adopted for the SD-750. 
In short, station-level triggers are first formed at each WCD. They are then combined with those from other detectors and examined for spatial and temporal correlations, leading to an array trigger, which initiates data acquisition. After that, a similar hierarchical selection of physics events out of the combinatorial background is ultimately made.

We describe in this section the design of the triggers (\cref{s:stationtriggers}).
We then illustrate their effect on the data, at the level of the amplitude of detected signals (\cref{s:additionalsignals}) and on the timing of detected signals in connection with the event selection (\cref{s:evselection}).
Finally we describe the energy at which acceptance is 100\% (\cref{s:efficiency}). 
A more detailed description of the trigger algorithms can be found in \ref{app::triggeralgo}.

\subsection{The Electromagnetic Triggers}
\label{s:stationtriggers} 

Using the station-level triggers, the digitized waveforms are constantly monitored in each detector for patterns consistent with what would be expected as a result of air-shower secondary particles (primarily electrons and photons of 10~MeV on average, and GeV muons) entering the water volume\footnote{The response of an individual WCD to secondary particles has been studied using unbiased FADC waveforms and dedicated studies of signals from muons~\cite{Bertou:2005ze}.}.
The typical morphologies include large signals, not necessarily spread in time, such as those close to the shower core, or sequences of small signals spread in time, such as those nearby the core in low-energy showers, or far from the core in high-energy ones.
Atmospheric muons, hitting the WCDs at a rate of 3\,kHz, are the primary background. 
The output from the PMTs has only a small dependence on the muon energy.
The electromagnetic and hadronic background, while also present, yields a total signal that is usually less than that of a muon.
Consequently, the atmospheric muons are the primary impediment to developing a station-level trigger for small signal sizes without contaminating the sampling of an air shower with spurious muons. 

Originally, two triggers were implemented into the station firmware, called \emph{threshold} (TH), more adept to detect muons, and \emph{time-over-threshold} (ToT), more suited to identify the electromagnetic component. Both of these have settings which require the signal to be higher in amplitude or longer than what is observed for a muon traveling vertically through the water volume.
As such, they have the inherent limitation of being insensitive to signals which are smaller than (or equal to) that of a single muon, thus prohibiting the measurement of pure electromagnetic signals, which are generally smaller. 

To bolster the sensitivity of the array to such small signals, two additional triggers were designed. The first, \emph{time-over-threshold-deconvolved} (ToTd), first removes the typical exponential decay created by
Cherenkov light inside the water volume, after which the ToT algorithm is applied. The second, {\it multiplicity-of-positive-steps} (MoPS), is designed to select small, non-smooth signals, a result of many electromagnetic particles entering the water over a longer period of time than a typical muon pulse. This is done by counting the number of instances in the waveform where consecutive bins are increasing in amplitude. Both of the trigger algorithms are described in detail in~\ref{app::triggeralgo}.

The implementation of the ToTd and MoPS (the rate of which is around 0.3\,Hz, compared to 0.6\,Hz of ToT and 20\,Hz of TH) did not require any modification in the logic of the array trigger, which calls for a coincidence of three or more SD stations that pass any combination of the triggers described above with compact spacing, spatially and temporally~\cite{Abraham:2010zz}. 
We note that in spite of the low rate of the ToTd and MoPS relative to TH and ToT, the array rate more than doubled after their implementation. 
This, as will be shown in the following, is due to the extension of measurements to the more abundant, smaller signals.

\subsection{Effect of ToTd and MoPS on Signals Amplitudes}
\label{s:additionalsignals}

The ToTd and MoPS triggers extend the range over which signals can be observed at individual stations into the region which is dominated by the background muons that are created in relatively low energy air showers.
By remaining insensitive to muon-like signals, these two triggers increase the sensitivity of the SD to the low-energy parts of the showers that have previously been below the trigger threshold.

\begin{figure}[t]
\centering
\includegraphics[width=0.99\columnwidth]{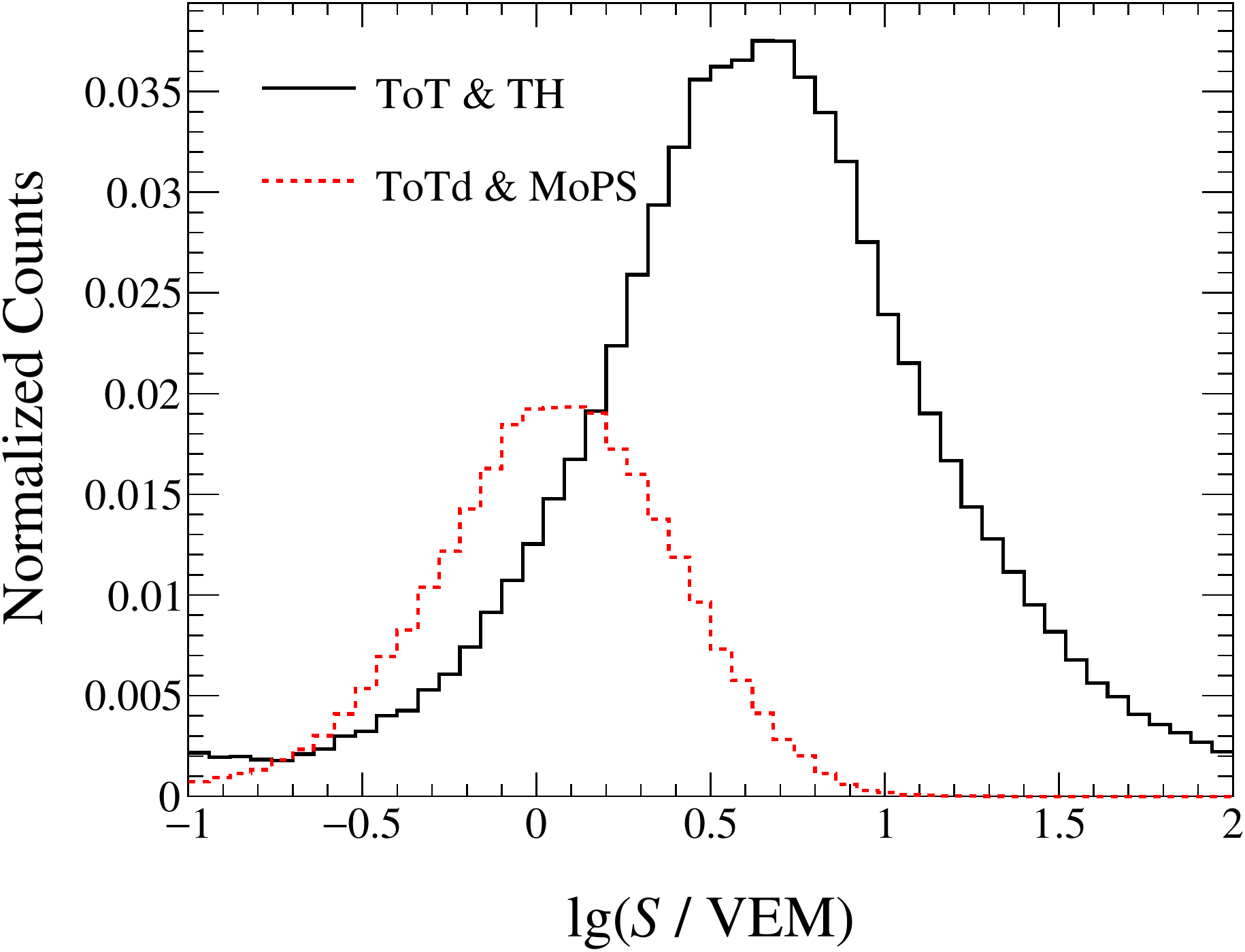}
\caption{Distribution of the signal sizes at individual stations which pass the TH and ToT triggers (solid black) and signals which pass \emph{only} the ToTd and/or MoPS triggers (dashed red).}
\label{f:SignalDistribution1}
\end{figure}

The effects of the additional triggers can be seen in the distribution of the observed signal sizes.
An example of such a distribution, based on one month of air-shower data, is shown in~\cref{f:SignalDistribution1}.
The signal sizes are shown in the calibration unit of one vertical equivalent muon (VEM), the total deposited charge of a muon traversing vertically through the water volume~\cite{Abraham:2010zz}.
For the stations passing only the ToT and TH triggers (shown in solid black), the distribution of deposited signals is the convolution of three effects,
the uniformity of the array, the decreasing density of particles as a function of perpendicular distance to the shower axis (henceforth referred to as the \emph{axial distance}), and the shape of the CR spectrum resulting in the negative slope above ${\simeq}7$\,VEM.
Furthermore there is a decreasing efficiency of the ToT and TH at small signal sizes.
The range of additional signals that are now detectable via the ToTd and MoPS triggers are shown in dashed red.
As expected, ToTd and MoPS triggers increase the probability of the SD to detect small amplitude signals, namely between 0.3 and 5\,VEM.
That the high-signal tail of this distribution ends near 10\,VEM is consistent with a previous study~\cite{ThePierreAuger:2015rma} that estimated that the ToT+TH triggers were fully efficient above this value.

\begin{figure}[t]
\centering
\includegraphics[width=0.99\columnwidth]{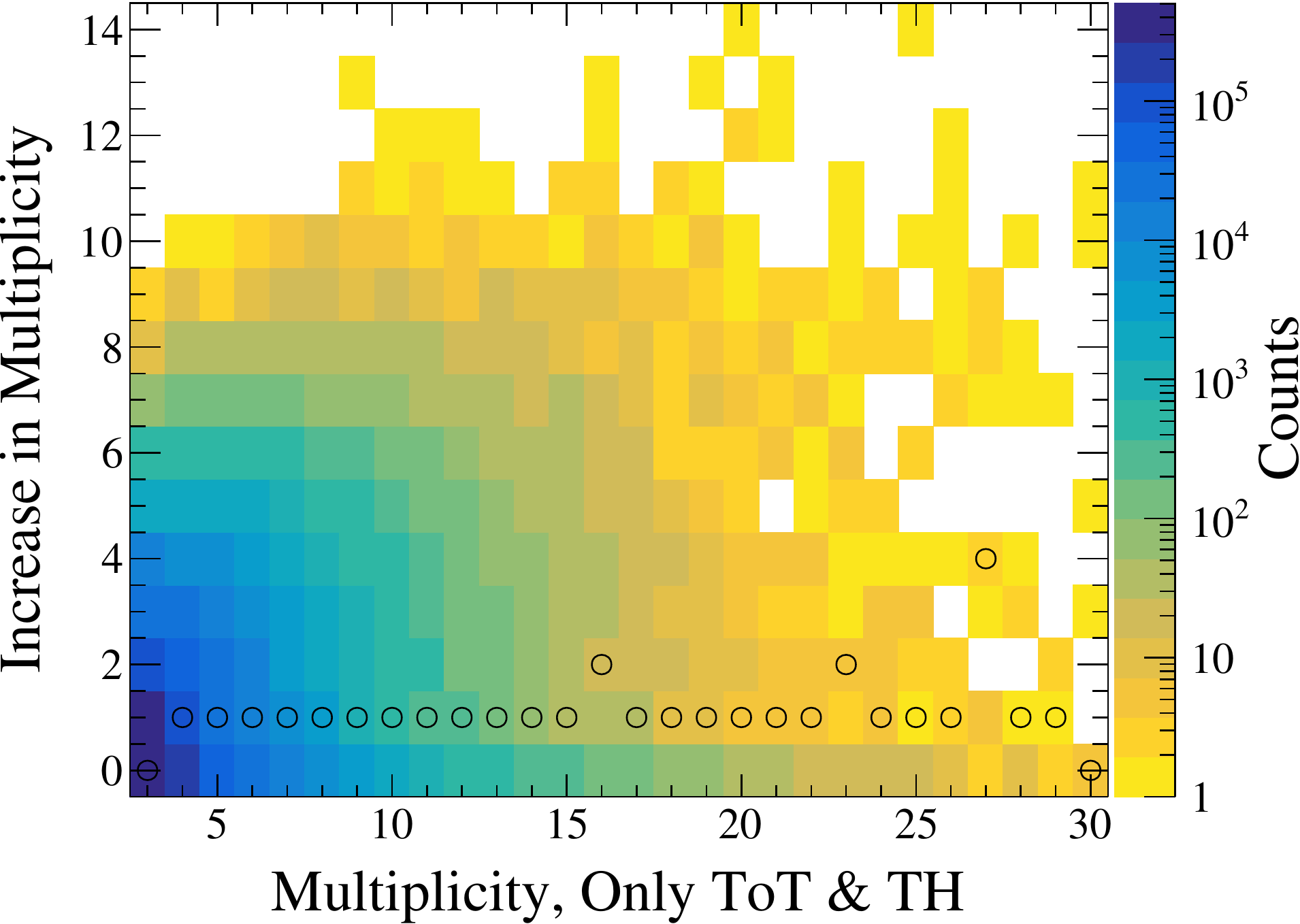}
\caption{The increase in station multiplicity when including the ToTd and MoPS triggers versus the original multiplicity with only ToT and TH. The black circles show the median increase in that multiplicity bin.}
\label{f:SignalDistribution2}
\end{figure}
The additional sensitivity to small air-shower signals also increases the multiplicity of triggered stations per event.
This increase is characterized in \cref{f:SignalDistribution2}, which shows the number of additional triggered stations per event as a function of the number of stations that pass the TH and ToT triggers, after removing spuriously triggered stations.
The median increase of multiplicity in each horizontal bin is shown by the black circles and indicates a typical increase of one station per event.

\subsection{Effects of ToTd and MoPS on Signal Timing}
\label{s:evselection}

The increased responsiveness of the ToTd and MoPS algorithms to smaller signals, specifically due to the electromagnetic component, has an effect also on the observed timing of the signals. In general, the electromagnetic signals are expected to be delayed with respect to the earliest part of the shower which is muon-rich, the delay increasing with axial distance. Further, in large events, stations that pass these triggers tend to be on the edge of the showers, where the front is thicker, thus increasing the variance of the arrival times. Such effects can be seen through the distribution of the start times for stations that pass the ToTd and MoPS triggers.

\begin{figure}[t]
\centering
\includegraphics[width=0.99\columnwidth]{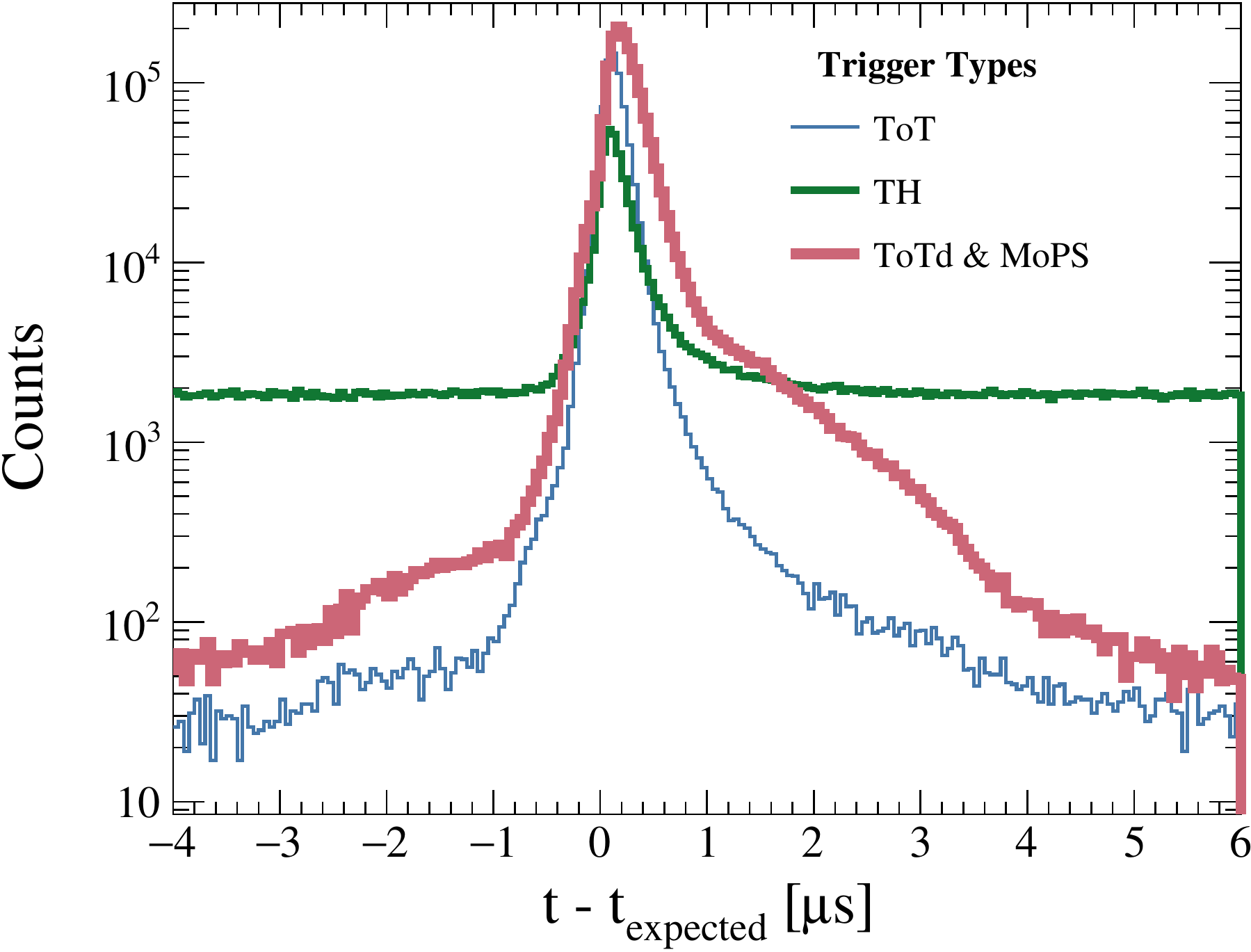}
\caption{Distributions of start times with respect to a plane front for stations that pass the ToT and TH algorithms, in blue and in green, respectively. The signals due to ToTd and MoPS are shown in red. Positive residuals correspond to a delay with respect to the plane wave expectation.}
\label{f:TimeResiduals}
\end{figure}

The residuals of the pulse start times with respect to a plane front fit of the three stations with the largest signals in the event are shown in Figure \ref{f:TimeResiduals} for different trigger types.
The entries shown in blue correspond to stations that passed the ToT algorithm, the ones in green to stations that pass the TH trigger (but not the ToT trigger), and those in red to stations that pass the ToTd and/or MoPS triggers, only.
For each of the trigger types, there is a clear peak near zero, which reflects the approximately planar shower front close to the core.
Stations that pass the TH condition, but not the ToT one, tend to capture isolated muons, including background muons arriving randomly in time.
This explains the vertical offset, flat and constant, in the green curve.
In turn, the lack of such a baseline shift in the blue and red distributions gives evidence that the ToT, TOTd and MoPS algorithms reject background muons effectively. 
This is particularly successful for the ToTd and MoPS that accept very small signals, of approximately 1\,VEM in size. 
One can see that these distributions have different shapes and that, in particular, the start time distributions of signals that pass the ToTd and MoPS have much longer tails than those of the TOT triggers, including a second distribution beginning around 1.5\,\textmu s possibly due to heavily delayed electromagnetic particles.

The extended time portion of showers accessed by the ToTd and MoPS triggers has implications on the procedure used to select physical events from the triggered ones~\cite{Abraham:2010zz}. 
In this process, non-accidental events, as well as
non-accidental stations, are disentangled on the basis of their timing.
First, we identify the combination of three stations where
they form a triangle, in which at least two legs are 750\,m long, and where they have the largest summed signal among all such possible configurations.
These stations make up the event \emph{seed} and the arrival times of the signals are fit to a plane front.
Additional stations are then kept if their temporal residual, $\Delta t$, is within a fixed window, $t_\text{low} < \Delta t < t_\text{high}$.
Motivated by the differing time distributions, updated $t_\text{low}$ and $t_\text{high}$ values were calculated based on which trigger algorithm was satisfied. Using the distributions of timing residuals, shown in \cref{f:TimeResiduals}, the baseline was first subtracted.
Then the limits of the window, $t_\text{low}$ and $t_\text{high}$, were chosen such that the middle 99\% of the distribution was kept.
The trigger-wise limits are summarized in \cref{t:SelectionLimits}.
\begin{table}
\caption{Temporal window limits $t_\text{low}$ and $t_\text{high}$ used to remove stations from an event, for each station-level trigger algorithm.}
\label{t:SelectionLimits}
\begin{tabular}{lll}
\hline\noalign{\smallskip}
\bf{Trigger Type} & \textbf{$\boldsymbol t_\text{low}$ [ns]} & \textbf{$\boldsymbol t_\text{high}$ [ns]}  \\
\noalign{\smallskip}\hline\noalign{\smallskip}
		ToT & $-397$ & 1454\\
		ToTd & $-468$ & 2285\\
		MoPS & $-477$ & 2883\\
		TH & $-485$ & 1379\\
\noalign{\smallskip}\hline
\end{tabular}
\end{table}

\subsection{Effect of the ToTd and MoPS on the energy above which acceptance is fully-efficient}
\label{s:efficiency}

Most relevant to the measurement of the spectrum is the determination of the energy threshold above which the SD-750 becomes fully efficient.
To derive this, events observed by the FD were used to characterize this quantity as a function of energy and zenith angle. 
The FD reconstruction requires only a single station be triggered to yield a robust determination of the shower trajectory.
Using the FD events with energies above $10^{16.8}$\,eV, the lateral trigger probability (LTP), the chance that a shower will produce a given SD trigger as a function of axial radius, was calculated for all trigger types.
The LTP was then parameterized as a function of the observed air-shower zenith angle and energy.
It is important to note that because the LTP is derived using observed air showers as a function of energy, this calculation reflects the efficiency as a function of energy based on the true underlying mass distribution of primary particles.
Further details of this method can be found in~\cite{auger2011LTP}.

\begin{figure}[t]
\centering
\includegraphics[width=0.99\columnwidth]{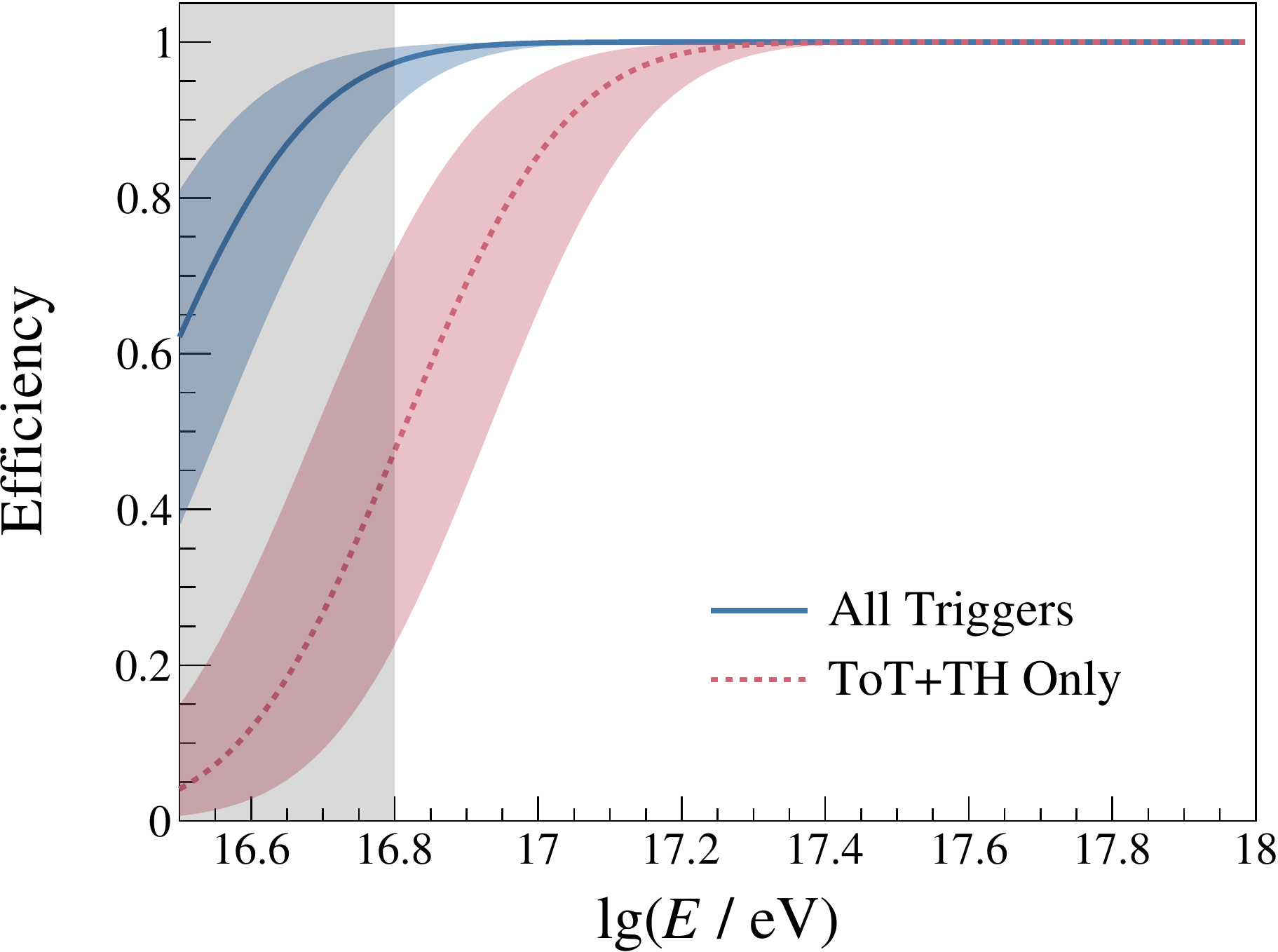}
\caption{The detection efficiency of the SD-750 for air showers with $\theta<40^\circ$ is shown for the original (dashed red) and expanded (solid blue) station-level trigger sets with bands indicating the systematic uncertainties. The trigger efficiency was determined using data above $10^{16.8}$\,eV and is extrapolated below this energy (shown in gray).}
\label{f:ArrayEfficiency}
\end{figure}

The SD-750 trigger efficiency was then determined via a study in which isotropic arrival directions and random core positions were simulated for fixed energies between $10^{16.5}$ and $10^{18}$\,eV.
Each station on the array was randomly triggered using the probability given by the LTP.
The set of stations that triggered were then checked against the compactness criteria of the array-level triggers, as described in~\cite{Abraham:2010zz}.
The resulting detection probability for showers with zenith angles $<40^\circ$ is shown as a solid blue line in \cref{f:ArrayEfficiency} as a function of energy. 
The detection efficiency becomes almost unity ($>98\%$) at around $10^{17}$~eV.\footnote{The energy-cut corresponding to the full-efficiency threshold increases with zenith angle, due to the increasing attenuation of the electromagnetic component with slant depth. The zenith angle $40^\circ$ was chosen as a balance to have good statistical precision and a low energy threshold.} For comparison, we show in the same figure, in dashed red, the detection efficiency curve for the original set of station-triggers, TH and ToT, in which the full efficiency is attained at a larger energy, i.e., around $10^{17.2}$~eV.

A description for the detection efficiency, $\epsilon(E)$, below $10^{17}$\,eV, will be important for unfolding the detector effects close to the threshold energy (see \cref{s:Spectrum}). 
This quantity was fit using the results of the LTP simulations with $\theta < 40^\circ$ and is well-parameterized by
\begin{equation}
    \begin{split}
    \epsilon(E) &= \frac{1}{2}\left[ 1 + \operatorname{erf}\left(\frac{\lg(E / \text{eV}) - \mu}{\sigma} \right) \right],
    \end{split}
    \label{eq:TriggerEff}
\end{equation}
where $\operatorname{erf}(x)$ is the error function, $\mu = 16.4 \pm 0.1$ and $\sigma = 0.261 \pm 0.007$.

For events used in this analysis, there is an additional requirement regarding the containment of the core within the array: only events in which the detector with the highest signal is surrounded by a hexagon of six stations that are fully operational are used. 
This criterion not only ensures adequate sampling of the shower but also allows the aperture of the SD-750 to be evaluated in a purely geometrical manner~\cite{Abraham:2010zz}. 
With these requirements, the SD-750 data set used below consists of about 560,000 events with $\theta < 40^\circ$ and $E>10^{17}$~eV recorded between 1 January 2014 and 31 August 2018.
The minimum energy cut is motivated by the lowest energy to which we can cross-calibrate with adequate statistics the energy scale of the SD with that of the FD (see \cref{s:ener_calib}).
The corresponding exposure, $\mathcal{E}$, after removal of time periods when the array was unstable\footnote{This is primarily due to the instabilities in the wireless communications systems as well as periods where large fractions of the array were not functioning.} (${<}2$\% of the total) is $\mathcal{E}=(105\pm 4)$\,km$^2$\,sr\,yr.
\section{Energy Measurements with the SD-750}
\label{s:EnergyMeasurements}

In this section, the method for the estimation of the air-shower energy is detailed together with the resulting energy resolution of the SD-750 array. The measurement of the actual shower size is first described in \cref{s:SizeEstimation} after which the corrections for attenuation effects are presented in \cref{s:Correction}. The energy calibration of the shower size after correction for attenuation is presented in \cref{s:ener_calib}. The energy resolution function is finally derived in \cref{s:res_func}.

\subsection{Estimation of the Shower Size}
\label{s:SizeEstimation}

The general strategy for the reconstruction of air showers using the SD-750 array is similar to that used for the SD-1500 array which is detailed extensively in~\cite{Aab:2020lhh}.
In this process, the arrival direction is obtained using the start times of signals, assuming either a plane or a curved shower front, as the degrees of freedom allow.
The lateral distribution of the signal is then fitted to an empirically-chosen function to infer the size of the air shower, which is used as a surrogate for the primary energy.
The reconstruction algorithm thus produces an estimate of the arrival direction and the size of the air shower via a log-likelihood minimization.

The lateral fall-off of the signal, $S(r)$, with increasing  distance, $r$, to the shower axis in the shower plane is modeled with a lateral distribution function (LDF). 
The stochastic variations in the location and character of the leading interaction in the atmosphere result in shower-to-shower fluctuations of the longitudinal development that propagate onto fluctuations of the lateral profile, sampled at a fixed depth. 
Showers induced by identical primaries at the same energy and at the same incoming angle can thus be sampled at the ground level at a different stage of development.
The LDF is consequently a quantity that varies on an event-by-event basis.
However, the limited degrees of freedom, as well as the sparse sampling of the air-shower particles reaching the ground, prevent the reconstruction of all the parameters of the LDF for individual events.
Instead, an \emph{average} LDF, $\langle S(r)\rangle$, is used in the reconstruction to infer the expected signal, $S(r_\text{opt})$, that would be detected by a station located at a reference distance from the shower axis, $r_\text{opt}$~\cite{Hillas1970,Newton2007}. 
This reference distance is chosen so as to minimize the fluctuations of the shower size, down to $\simeq 7\%$ in our case. 
The observed distribution of signals is then adjusted to $\langle S(r)\rangle$ by scaling the normalization, $S(r_\text{opt})$, in the fitting procedure.

%
%
%
%
The reference distance, or optimal distance, $r_\text{opt}$, has been determined on an event-by-event basis by fitting the measured signals to different hypotheses for the fall-off of the LDF with distance to the core as in~\cite{Newton2007}. 
Via a fit of many power-law-like functions, the dispersion of signal expectations has been observed to be minimal at $r_\text{opt}\simeq 450$\,m, which is primarily constrained by the geometry of the array. The expected signal at 450\,m from the core, $S(450)$, has thus been chosen to define the shower-size estimate.  

The functional shape chosen for the average LDF is a parabola in a log-log representation of $\langle S(r)\rangle$ as a function of the distance to the shower core,
\begin{equation}
\ln \langle S(r) \rangle= \ln S(450)+\beta\,\rho + \gamma\,\rho^2,
\label{eq:loglog}
\end{equation}
where $\rho=\ln(r/(450\,\text{m}))$, and $\beta$ and $\gamma$ are two structure parameters. 
The overall steepness of the fall-off of the signal from the core is governed by $\beta$, while the concave deviation from a power-law function is given by $\gamma$. 
The values of $\beta$ and $\gamma$ have been obtained in a data-driven manner, by using a set of air-shower events with more than three stations, none of which have a saturated signal.
The zenith angle and the shower size are used to trace the age dependence of the structure parameters based on the following parameterization in terms of the reduced variables $t=\sec\theta - 1.27$ and $u=\ln S(450) - 5$:
\begin{eqnarray}
\beta &=& (\beta_0 + \beta_1 t + \beta_2 t^2)(1 + \beta_3 u),\\
\gamma &=& \gamma_0 + \gamma_1 u.
\label{eq:SlopeParameters}
\end{eqnarray}
For any specific set of values $\mathbf{p}=\{\beta_i, \gamma_i\}$, the reconstruction is then applied to calculate the following $\chi^2$-like quantity, globally to all events:
\begin{equation}
    Q^2(\mathbf{p})=\frac{1}{N_\text{tot}}\sum_{k=1}^{N_\text{events}}\sum_{j=1}^{N_k}\frac{(S_{k,j}-\langle S(r_j,\mathbf{p})\rangle)^2}{\sigma_{k,j}^2}.
\end{equation}
The sum over $N_k$ stations is restricted to those with observed signals larger than 5\,VEM to minimize the impact of upward fluctuations of the station signals far from the core and hence to avoid biases from trigger effects, and to stations more than 150\,m away from the core. 
The uncertainty $\sigma_{k,j}$ is proportional to $\sqrt{S_{k,j}}$~\cite{Aab:2020lhh}. 
$N_\text{tot}$ is the total number of stations in all such events.
The best-fit \{$\beta_i$, $\gamma_i$\} values are collected in \cref{tab:structure_parameters}. 

\begin{table}
\caption{Best-fit \{$\beta_i$, $\gamma_i$\} values defining the structure parameters of the LDF.}
\label{tab:structure_parameters}       
\begin{tabular}{l S[table-format=-1.2(1), separate-uncertainty=true]}
\hline\noalign{\smallskip}
\textbf{Parameter} & \textbf{Value}  \\
\noalign{\smallskip}\hline\noalign{\smallskip}
		$\beta_0$ & 2.95(2)\\
		$\beta_1$ & -1.0(2)\\
		$\beta_2$ & 0.7(2)\\
		$\beta_3$ & 0.02(1)\\
		$\gamma_0$ & 0.26(9)\\
		$\gamma_1$ & -0.02(1)\\
\noalign{\smallskip}\hline
\end{tabular}
\end{table}

\subsection{Correction of Attenuation Effects}
\label{s:Correction}

There are two significant observational effects that impact the precision of the estimation of the shower size.
Both of these effects are primarily a result of the variable slant depth that a shower must traverse before being detected with the SD. Since the mean atmospheric overburden is 875\,g/cm$^2$ at the location of the Observatory,  nearly all observed showers in the energy range considered in this analysis have already reached their maximum size and have started to attenuate~\cite{bellido2018depth}.
Thus, an increase in the slant depth of a shower results in a more attenuated cascade at the ground, directly impacting the observed shower size.

The first observational effect is related to the changing weather at the Observatory.
Fluctuations in the air pressure equate to changes in the local overburden and thus showers observed during periods of relatively high pressure result in an underestimated shower size.
Similarly, the variations in the air density directly change the Moli\`ere radius which directly affects the spread of the shower particles.
The increased lateral spread of the secondaries, or equivalently, the decrease in the density of particles on the ground, also leads to a systematically underestimated shower size.
Both the air-density and pressure have typical daily and yearly cycles that imprint similar cycles upon the estimation of the shower size. 

The relationship between these two atmospheric parameters and the estimated shower sizes has been studied using events detected with the SD~\cite{aab:2017impact}.
From this relationship, a model was constructed to scale the observed value of $S(450)$ to what would have been measured had the shower been instead observed at a time with the daily and yearly average atmosphere.
When applying this correction to individual air showers, the measurements from the weather stations located at the FD sites are used.
The values of $S(450)$ are scaled up or down according to these measurements, resulting in a shift of at most a few percent. 
The shower size is eventually the proxy of the air-shower energy, which is calibrated with events detected with the FD (see \cref{s:ener_calib}). Since the FD operates only at night when, in particular, the air density is relatively low, the scaling of $S(450)$ to a daily and yearly average atmosphere corrects for a ${\simeq}0.5\%$ shift in the assigned energies.

The second observational effect is geometric, wherein showers arriving at larger zenith angles have to go through more atmosphere before reaching the SD.
To correct for this effect, the Constant Intensity Cut (CIC) method~\cite{Hersil:1961zz} is used.
The CIC method relies on the assumption that cosmic rays arrive isotropically, which is consistent with observations in the  energy range considered~\cite{Aab:2020xgf}.
The intensity is thus expected to be independent of arrival direction after correcting for the attenuation.
Deviations from a constant behavior can thus be interpreted as being due to attenuation alone.
Based on this property, the CIC method allows us to determine the attenuation curve as function of the zenith angle and therefore to infer a zenith-independent shower-size estimator.

We empirically chose a functional form which describes the relative amount of attenuation of the air shower,
\begin{equation}
    f_\text{CIC}(\theta) = 1 + a x + bx^2.
\label{eq:cic}
\end{equation}
The scaling of this function is normalized to the attenuation of a shower arriving at $35^\circ$ by choosing $x = \sin^2 35^\circ - \sin^2 \theta$.
For a given air shower, the observed shower size can be scaled using \cref{eq:cic} to get the equivalent signal of a shower arriving with the reference zenith angle, $S_{35}$, via the relationship $S(450) = S_{35}\,f_\text{CIC}(\theta)$.

Isotropy implies that ${\mathrm{d}N/\mathrm{d}\sin^2\theta}$ is constant.
Thus, the shape of $f_\text{CIC}(\theta)$ is determined by finding the parameters $a$ and $b$ for which the CDF of events above $S(450) > S_\text{cut}\, f_\text{CIC}(\theta)$ is linear in $\sin^2 \theta$ using an Anderson-Darling test~\cite{anderson1954darling}.
The parameter $S_\text{cut}$ defines the size of a shower with $\theta = 35^\circ$ at which the CIC tuning is performed, the choice of which is described below.

Since the attenuation that a shower undergoes before being detected is related to the depth of shower maximum and the particle content, the shape of $f_\text{CIC}(\theta)$ is dependent on both the energy and the average mass of the primary particles at that energy.
Further, this implies that a single choice of $S_\text{cut}$ could introduce a mass and/or energy bias.
Thus, \cref{eq:cic} was extended to allow the polynomial coefficients, $k \in \{a,\,b\}$, to be functions of $S(450)$ via $k( S(450)) = k_0 + k_1 y + k_2 y^2$ where $y = \lg (S(450) / \text{VEM})$.
The function $f_\text{CIC}(\theta, S(450))$ was tuned using an unbinned likelihood.

\begin{figure}[t]
\centering
\includegraphics[width=0.99\columnwidth]{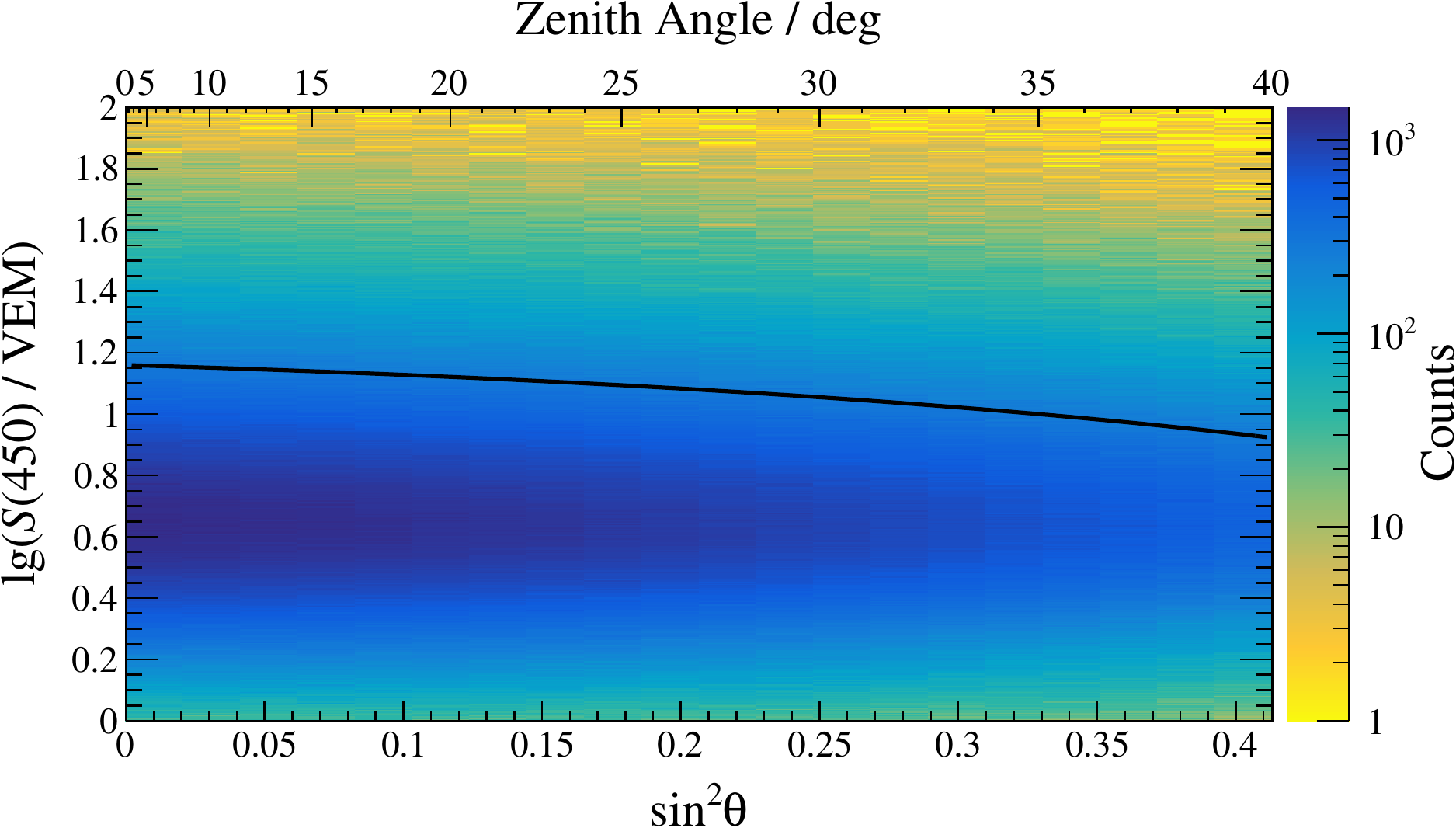}\\
\vspace{0.3cm}
\includegraphics[width=0.99\columnwidth]{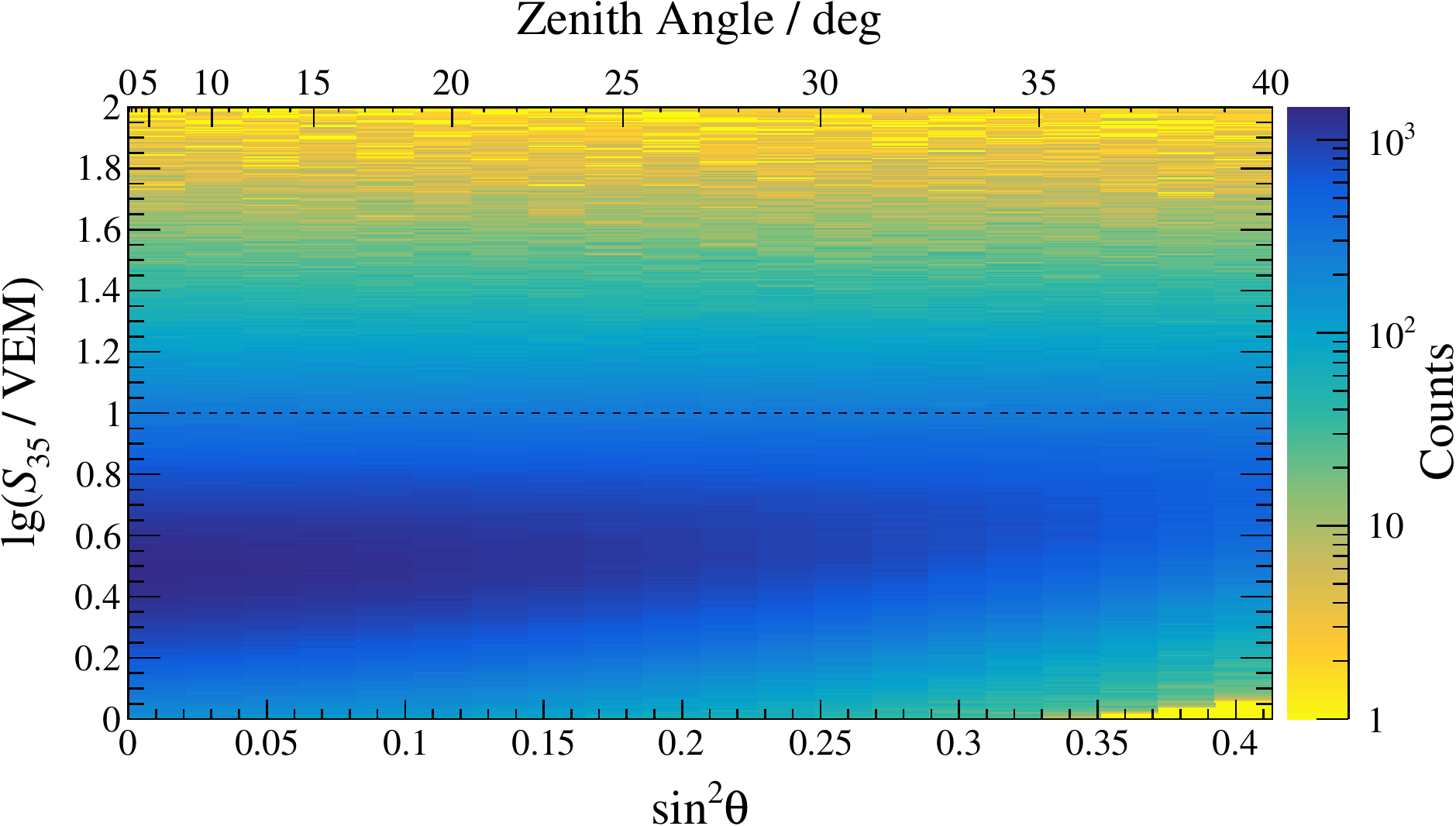}
\caption{Top: histogram of reconstructed shower sizes and zenith angles. The solid black line represents the shape of $f_\text{CIC}$ at 10\,VEM. Bottom: same distribution but as a function of corrected shower size, $S_{35}$, and zenith angle. The dashed black line indicates the mapping of the solid black line in the top figure after inverting the effects of the CIC correction.}
\label{f:CIC}
\end{figure}

The fit was performed so as to guarantee equal intensity of the integral spectra using eight threshold values of $S_\text{cut}$ between 10 and 70\,VEM, evenly spaced in log-scale.
These values were chosen to avoid triggering biases on the low end and the dwindling statistics on the high end.
The best fit parameters are given in \cref{t:CICCoeff}.
The resulting 2D distribution of the number of events, in equal bins of $\sin^2\theta$ and $\lg S_{35}$, is shown in \cref{f:CIC}, bottom panel.
It is apparent that the number of events above any $\sin^2{\theta}$ value is equalized for any constant line for $\lg S_{35}\gtrsim 0.7$.
The magnitude of the CIC correction is $(-27\pm4)$\% for vertical showers (depending on $S(450)$) and $+15$\% for a zenith angle of $40^\circ$.

\begin{table}
\caption{The energy dependence of the CIC parameters (\cref{eq:cic}) are given below. }
\label{t:CICCoeff}       
\begin{tabular}{llll}
\hline\noalign{\smallskip}
& \multicolumn{1}{c}{$k_0$} & \multicolumn{1}{c}{$k_1$} & \multicolumn{1}{c}{$k_2$}  \\
\noalign{\smallskip}\hline\noalign{\smallskip}
		$a$ & $\phantom{-}2.42$ & $-0.886$ & $\phantom{-}0.268$\\	
		$b$ & $-4.56$ & $\phantom{-}5.61$ & $-2.47$\\
\noalign{\smallskip}\hline
\end{tabular}
\end{table}

\subsection{Energy Calibration of the Shower Size}
\label{s:ener_calib}

\begin{figure}[h]
\centering
\includegraphics[width=0.99\columnwidth]{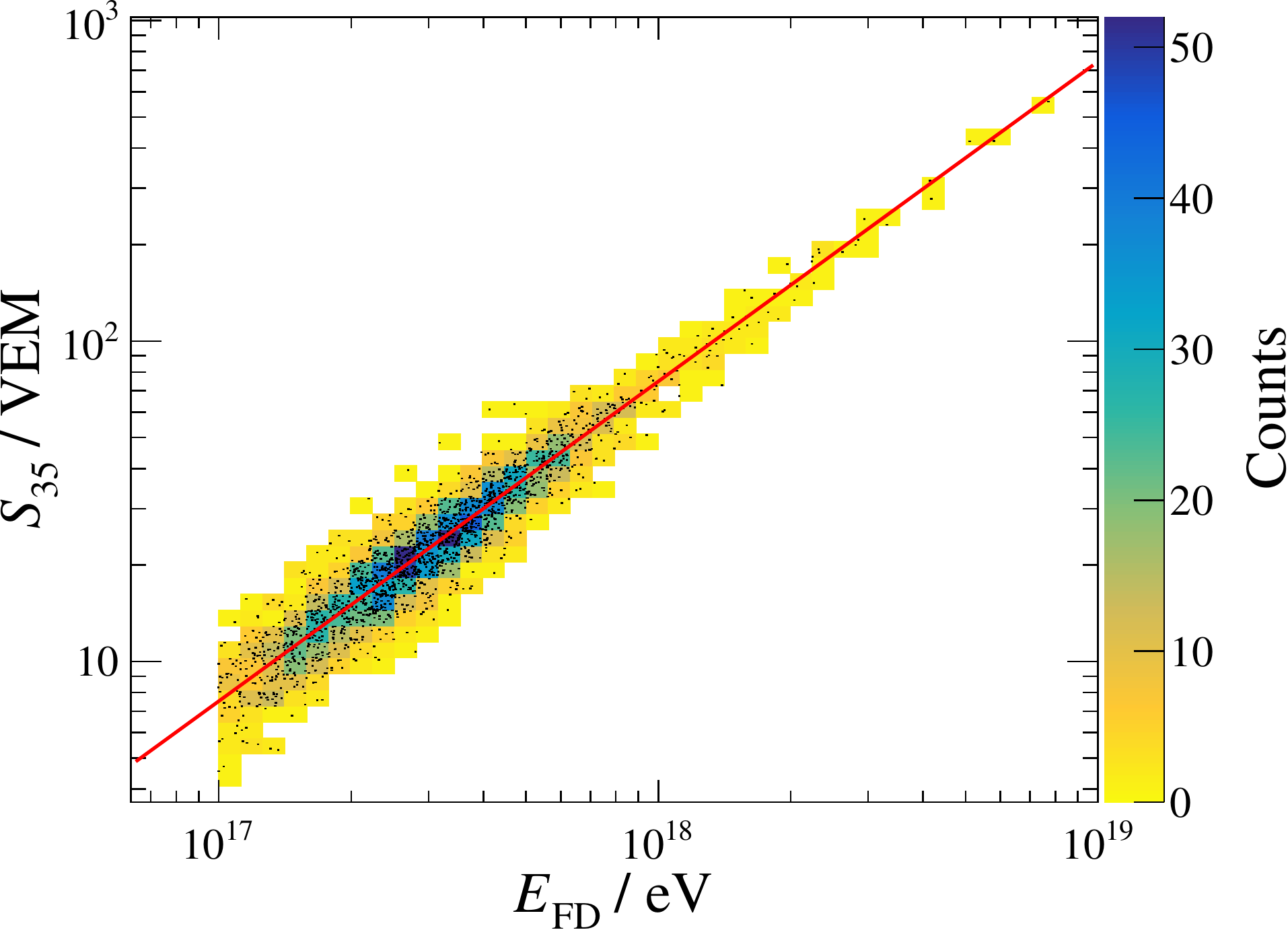}
\caption{Correlation between the SD shower-size estimator, $S_{35}$, and the reconstructed FD energy, $E_\text{FD}$, for the selected hybrid events.}
\label{f:EnergyCalib}
\end{figure}
The conversion of the shower size, corrected for attenuation, is based on a special set of showers, called \emph{golden hybrid} events, which can be reconstructed independently by the FD and by the SD. 
The FD allows for a calorimetric estimate of the primary energy except for the contribution carried away by particles that reach the ground. 
The amount of this so-called \emph{invisible energy}, ${\simeq}20\%$ at $10^{17}$\,eV and ${\simeq}15\%$ at $10^{18}$\,eV, has been evaluated using simulations~\cite{Aab:2019cwj} tuned to measurements at $10^{18.3}$\,eV so as to correct for the discrepancy in the muon content of simulated and observed showers~\cite{Aab:2020frk}. 
The empirical relationship between the FD energy measurements, $E_\text{FD}$, and the corrected SD shower size, $S_{35}$, allows for the propagation of the FD energy scale to the SD events. 

FD events were selected based on quality and fiducial criteria aimed at guaranteeing a precise estimation of $E_\text{FD}$ as well as at minimizing any acceptance biases towards light or heavy mass primaries introduced by the field of view of the FD telescopes.
The cuts used for the energy calibration are similar to those described in~\cite{Aab:2014aea,bellido2018depth}. They include the selection of data when the detectors are properly operational and the atmosphere properties like clouds coverage and the vertical aerosol depth are suitable for a good determination of the air-shower profile.  A further quality selection includes requirements on the uncertainties of the energy assignment (less than 12\%) and of the reconstruction of the depth at the maximum of the air-shower development (less than 40~g~cm$^{-2}$). A possible bias due to a selection dependency on the primary mass is avoided by using an energy dependent fiducial volume determined from data as in~\cite{bellido2018depth}.


Restricting the data set to events with $E_\text{FD} \geq 10^{17}$\,eV, {\it (to ensure that the SD is operating in the regime of full efficiency)} there are 1980 golden-hybrid events available to establish the relationship between $S_{35}$ and $E_\text{FD}$. Fourty-five events in the energy range between $10^{16.5}$\,eV and $10^{17}$\,eV are included in the likelihood as described in~\cite{Dembinski:2015wqa}.
As $S_{35}$ depends on the mass composition of the primary particles, the relation between $S_{35}$ and $E_\text{FD}$, shown in \cref{f:EnergyCalib}, accounts for the trend of the composition change with energy inherently as the underlying mass distribution is directly sampled by the FD.
Measurements of $\langle X_\text{max}\rangle$ suggest that this composition trend follows a logarithmic evolution up to an energy of $10^{18.3}$\,eV, beyond which the number of events available for this analysis is too small to affect the results in any way~\cite{Aab:2014aea}.
So we choose a power-law type relationship,
\begin{equation}
    E_{\rm SD}=A S_{35}^B,
    \label{eq:ecalib}
\end{equation}
which is expected from Monte-Carlo simulations in the case of a single logarithmic dependence of $X_\text{max}$ with energy.
The energy of an event with $S_{35} = 1$\,VEM arriving at the reference angle, $A$, and the logarithmic slope, $B$, are fitted to the data by means of a maximum likelihood method which models the distribution of golden-hybrid events in the plane of energies and shower sizes. 
The use of these events allows us to infer $A$ and $B$ while accounting for the clustering of events in the range $10^{17.4}$ to $10^{17.7}$\,eV observed in \cref{f:EnergyCalib} due to the fall-off of the energy spectrum combined with the restrictive golden-hybrid acceptance for low-energy, dim showers. 
A comprehensive derivation of the likelihood function can be found in~\cite{Dembinski:2015wqa}. 

The probability density function entering the likelihood procedure, detailed in~\cite{Dembinski:2015wqa}, is built by folding the cosmic-ray intensity, as observed through the effective aperture of the FD, with the resolution functions of the FD and of the SD. 
Note that to avoid the need to model accurately the cosmic-ray intensity observed through the effective aperture of the telescopes (and thus to reduce reliance on mass assumptions), the observed distribution of events passing the cuts described above is used. 
The FD energy resolution, $\sigma_\text{FD}(E)/E_\text{FD}$, is typically between 6\% and 8\%~\cite{Dawson:2020bkp}.
It results from the statistical uncertainty arising from the fit to the longitudinal profile, the uncertainties in the detector response, the uncertainties in the models of the state of the atmosphere, and the uncertainties in the expected fluctuations from the invisible energy. 
The SD shower-size resolution, $\sigma_\text{SD}(S_{35})/S_{35}$, is, on the other hand, comprised of two terms, the detector sampling fluctuations, $\sigma_\text{det}(S_{35})$, and the shower-to-shower fluctuations, $\sigma_\text{sh}(S_{35})$.
The former is obtained from the sum of the squares of the uncertainties from the reconstructed shower size and zenith angle, and from the
attenuation-correction terms that make up the $S_{35}$ assignment. 
The latter stem from the stochastic nature of both the depth of first interaction of the primary and the subsequent development of the particle cascade.  
This contribution thus depends on the CR mass composition and on the hadronic interactions in air showers. 
For this reason, the derivation of $A$ and $B$ follows a two-step procedure. 
A first iteration of the fit is carried out by using an educated guess for $\sigma_\text{sh}(S_{35})$, as expected from Monte-Carlo simulations for a mass-composition scenario compatible with data~\cite{bellido2018depth}.
The total resolution $\sigma_\text{SD}(S_{35})/S_{35}$ is then extracted from data as explained next in \cref{s:res_func} and used in a second iteration.

\begin{table}
\caption{The systematic uncertainties on the FD energy scale are given below. Lines with multiple entries represent the values at the low and high end of the considered energy range ($\simeq$10$^{17}$ and $\simeq$10$^{19}$\,eV, respectively).}
\label{t:EFDSys}       
\begin{tabular}{ll}
\hline\noalign{\smallskip}
Systematic & Uncertainty  \\
\noalign{\smallskip}\hline\noalign{\smallskip}
        Absolute fluorescence yield & 3.6\% \\
        Atmosphere and scattering & 2 to 6\% \\
        FD Calibration & 10\% \\
        Longitudinal profile reconstruction & 7 to 5.5\% \\
        Invisible energy & 3 to 1.5\% \\
\noalign{\smallskip}\hline
\end{tabular}
\end{table}

The resulting relationship is shown as the red line in \cref{f:EnergyCalib} with best-fit parameters such that $A=(13.2\pm0.3)$\,PeV and $B=1.002\pm 0.006$. 
The goodness of the fit is supported by the $\chi^2/\text{NDOF} = 2120/1978$ ($p = 0.013$). 
We use these values of $A$ and $B$ to calibrate the shower sizes in terms of energies by defining the SD estimator of energies, $E_\text{SD}$, according to \cref{eq:ecalib}.
The SD energy scale is set by the calibration procedure and thus it inherits the $A$ and $B$ calibration-parameters uncertainties and the FD energy-scale uncertainties, listed in \cref{t:EFDSys}.
The systematic uncertainty, after addition in quadrature, of the energy scale is about 14\% and is almost energy independent.
The energy independence is a consequence of the 10\% uncertainty of the FD calibration, which is the dominant contribution.

\subsection{Resolution Function of the SD-750 Array}
\label{s:res_func}

The SD resolution as a function of energy is needed in several steps of the analysis. 
In the regime of full efficiency, it can be considered as a Gaussian function centered on the true energy, the width of which reflects the statistical uncertainty associated with the detection and reconstruction processes on one hand, and the stochastic development of the particle cascade on the other hand.
The combination of the two can be estimated for the golden hybrid events, thus allowing us to account for the contribution of the shower-to-shower fluctuations in a data-driven way.

Each event observed by the SD and FD results in two independent measurements of the air-shower energy, $E_\text{SD}$ and $E_\text{FD}$, respectively.
Unlike for the SD, the FD directly provides a view of the shower development so a total energy resolution, $\sigma_\text{FD}(E)$, can be estimated for each of the golden hybrid events.
Using the known $\sigma_\text{FD}(E)$, the resolution of SD can be determined by studying the distribution of the ratio of the two energy measurements.

\begin{figure}[t]
\centering
\includegraphics[width=0.99\columnwidth]{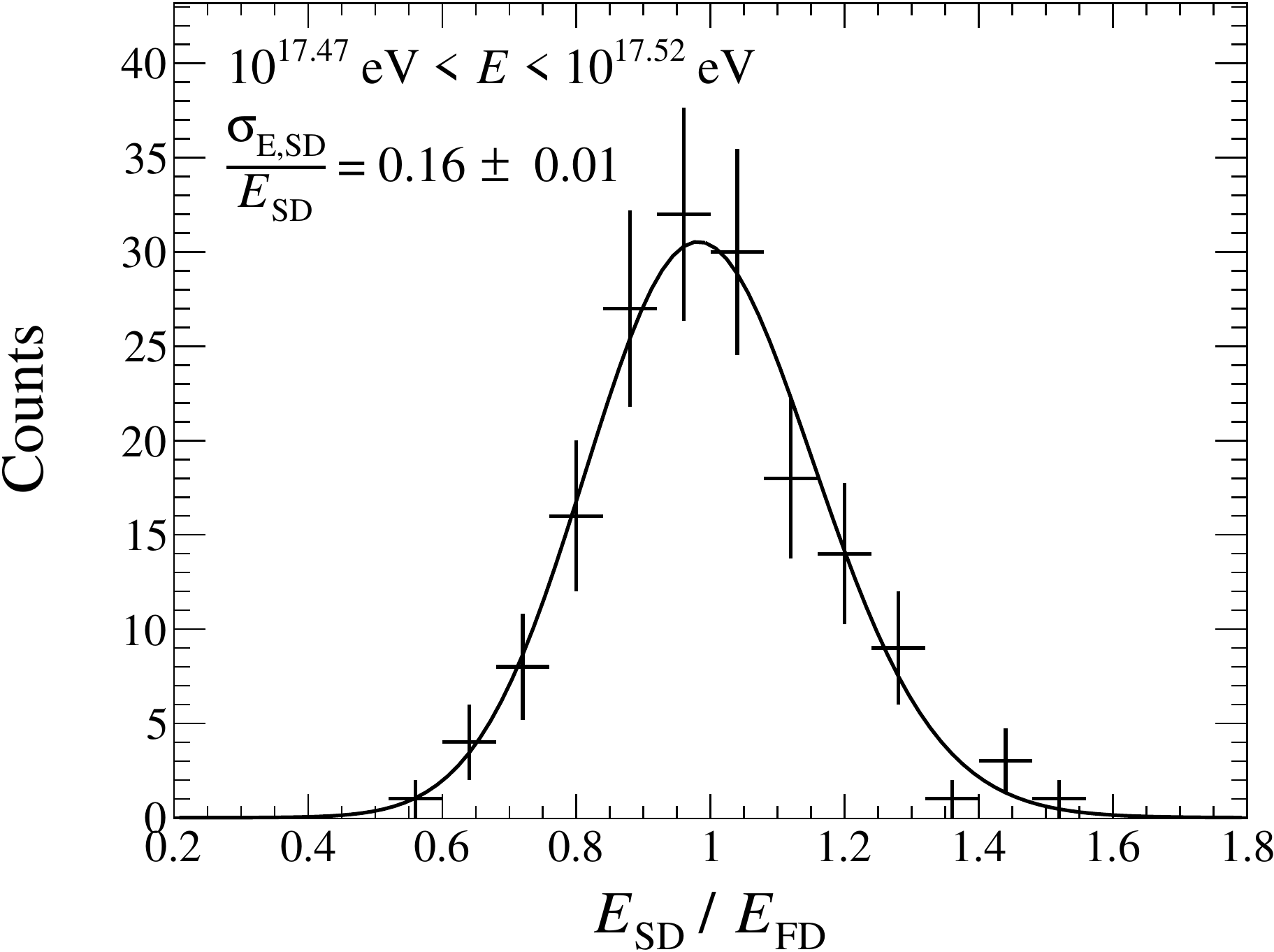}
\caption{An example of the ratio of the energy assignments for the SD and FD is shown with black crosses for the energy bin indicated in the plot. The best fit ratio distribution for this bin is shown by the black line.}
\label{f:EnergyRatio}
\end{figure}

For two independent, Gaussian-distributed random variables, $X$ and $Y$, their ratio, $z=X/Y$, produces a \emph{ratio distribution} that depends on the means ($\mu_X$, $\mu_Y$) and standard deviations ($\sigma_X$, $\sigma_Y$) of the two variables, $\operatorname{PDF}(z; \mu_X, \mu_Y, \sigma_X, \sigma_Y)$.
Likewise, the ratio of the two energy measurements, $z = E_\text{SD} / E_\text{FD}$, follows such a distribution to first order.
Because the FD sets the energy scale of the Observatory, there is inherently no bias in the energy measurements with respect to its own scale and thus, on average, $\mu_\text{FD}(E)=1$.
Using the golden hybrid data set, the ratio distribution was fit in an unbinned likelihood analysis,
$\operatorname{PDF}(z; \mu_\text{SD}(E), 1, \sigma_\text{SD}(E), \sigma_\text{FD}(E))$.

\begin{figure}[t]
\centering
\includegraphics[width=0.99\columnwidth]{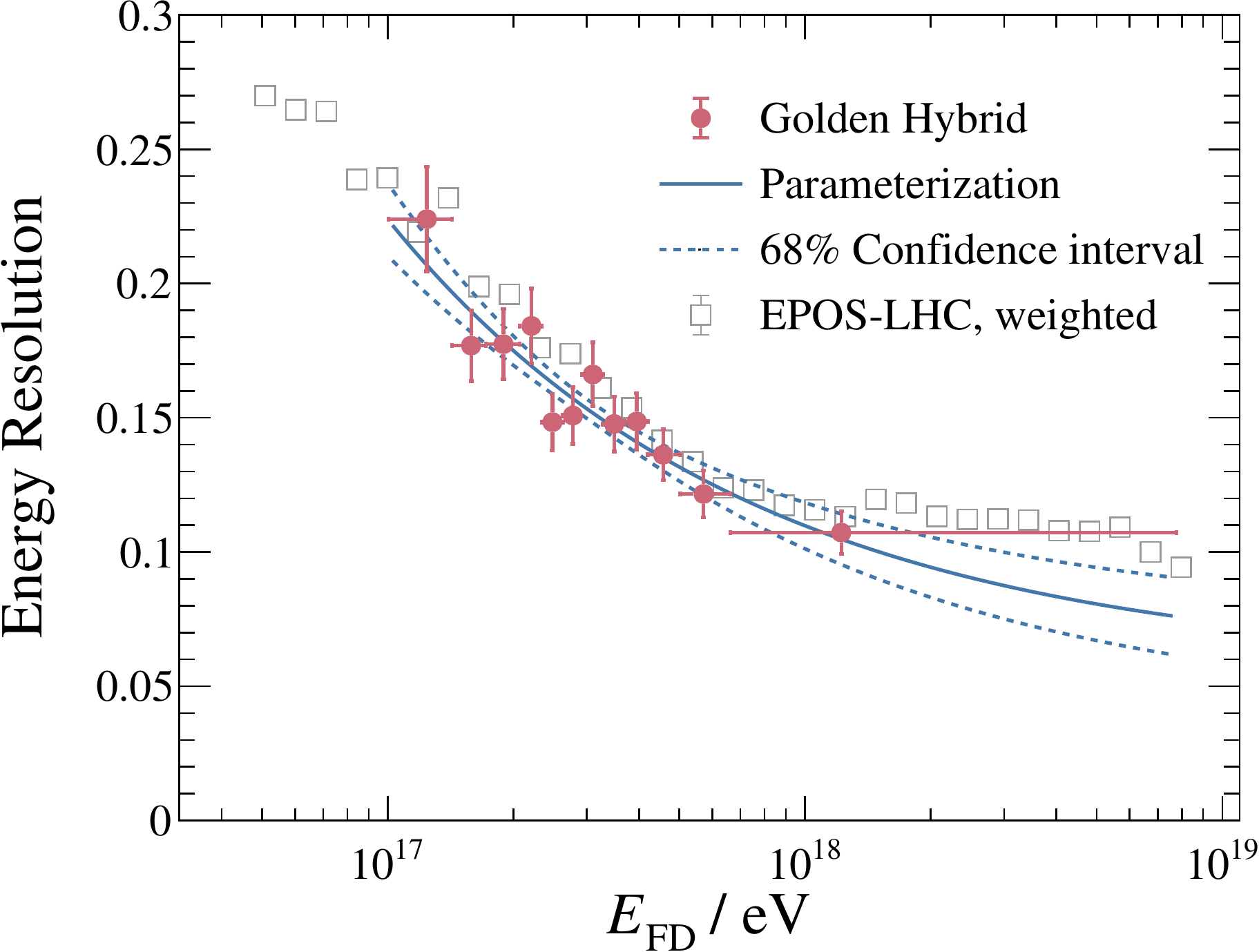}
\caption{The total SD energy resolution, as calculated using the golden hybrid events (red circles) is shown in bins with equal statistics. The parameterization of the resolution is shown by the solid blue line and the corresponding 68\% confidence interval in dashed lines. The energy resolution, calculated using mass-weighted MC air showers (gray squares), is shown as a verification of the method.}
\label{f:EnergyResolution}
\end{figure}

An example of the measured energy-ratio distributions is shown in \cref{f:EnergyRatio} with the fitted curve overlaid on the data points.
Carrying out the fit in different energy bins, the SD resolution, shown by the red points in \cref{f:EnergyResolution}, is represented by,
\begin{equation}
    \frac{\sigma_\text{SD}(E)}{E} = (0.06 \pm 0.02) + (0.05 \pm 0.01) \sqrt{\frac{1\,\text{EeV}}{E}}.
    \label{eq:EnergyResolution}
\end{equation}
The corresponding curve is overlaid in blue, bracketed by the 68\% confidence region.

To measure the spectrum above the $10^{17}$\,eV threshold, the knowledge of the resolution function, which induces bin-to-bin migration of events, and of the detection efficiency are also required for energies below this threshold.
As a verification, particularly in the energy region where \cref{eq:EnergyResolution} is extrapolated, a Monte-Carlo analysis was performed.
A set of 325,000 CORSIKA~\cite{heck1998corsika} air showers were used, consisting of proton, helium, oxygen, and iron primaries with energies above $10^{16}$\,eV.
EPOS-LHC~\cite{Pierog:2013ria} was used as the hadronic interaction model.
The air showers were run through the full SD simulation and reconstruction algorithms.
The events were weighted based on the primary mass according to the Global Spline Fit (GSF) model~\cite{dembinski2018gsf} to account for the changing mass-evolution near the second knee and ankle.
The reconstructed values of $S(450)$ were corrected by applying the energy-dependent CIC method to obtain values for $S_{35}$ and these values were then calibrated against the Monte-Carlo energies.
During the calibration, a further weighting was performed based on the energy distribution of golden hybrid events to account for the hybrid detection efficiency.
Following the calibration procedure, each MC event was assigned an energy in the FD energy scale (i.e.\ $E_\text{MC} \to S_{35} \to E_\text{FD}$).

The SD energy resolution was calculated using the mass-weighted simulations and is shown in gray squares in \cref{f:EnergyResolution}.
Indeed, the simulated and measured SD resolutions show a similar trend and agree to within the uncertainties, supporting the golden hybrid method.

In the energy region at-and-below $10^{17}$\,eV, systematic effects also enter into play on the energy estimate. 
An energy-dependent offset, a bias, is thus expected in the resolution function for several reasons:
\begin{enumerate}
    \item The application of the trigger below threshold, combined with the finite energy resolution, cause an overestimate of the shower size, on average, which is then propagated to the energy assignment.
    \item The linear relationship assumed in \cref{eq:ecalib} cannot account for a possible sudden change in the evolution of the mass-composition with energy. Such a change would require a broken power law for the energy calibration relationship.
    \item In the energy range where the SD is not fully efficient, the SD efficiency is larger for light primary nuclei, thus preventing a fair sampling of $S_{35}$ values over the underlying mass distribution.
\end{enumerate}

\begin{figure}[t]
\centering
\includegraphics[width=0.99\columnwidth]{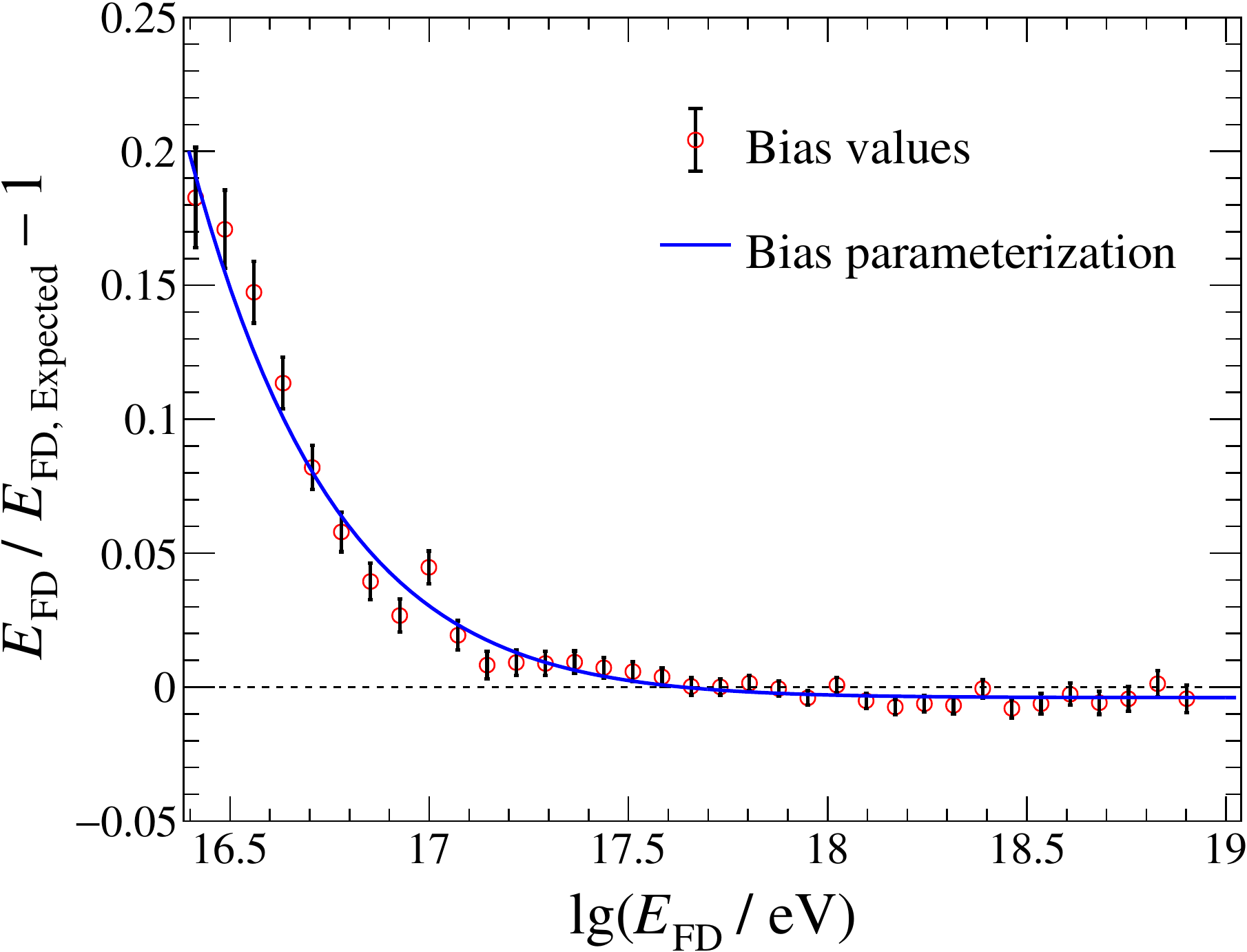}
\caption{The bias of the energy assignment for the SD-750 was studied using Monte Carlo simulations, weighted according to the GSF model~\cite{dembinski2018gsf}. The ratio of the assigned and expected values as a function of energy are shown (red circles) along with the parameterization (blue line) given in \cref{eq:EnergyBias}.}
\label{f:EnergyBias}
\end{figure}

Because there is an insufficient number of FD events which pass the fiducial cuts below $10^{17}$\,eV, the bias was characterized, using the same air-shower simulations as used for the resolution cross-check.
The remaining relative energy bias is shown in \cref{f:EnergyBias}.
The ratio between the reconstructed and expected values are shown as the red points as a function of $E_\text{FD}$.
A larger bias of $\simeq$20\% is seen at low energies, where upward fluctuations are necessarily selected by the triggering conditions.
In the range considered for the energy spectrum, $E > 10^{17}$\,eV, the bias is 3\% or less.
To complete the description of the SD resolution function, the relative bias was fit to an empirical function,
\begin{equation}
    b_\text{SD}(E)= b_0 (\lg\tfrac{E}{\mathrm{eV}} - b_1)\exp \left(-b_2(\lg\tfrac{E}{\mathrm{eV}} - b_3)^2\right) + b_4.
    \label{eq:EnergyBias}
\end{equation}
The corresponding best fit parameters (blue line in \cref{f:EnergyBias}) are given in \cref{tab:bias_parameters}.
\begin{table}
\caption{Best-fit parameters for the relative energy bias of the SD-750, $ b_\text{SD}(E)$, given in \cref{eq:EnergyBias}.}
\label{tab:bias_parameters}
\begin{tabular}{lll}
\hline\noalign{\smallskip}
\textbf{Parameter} & \textbf{Value} & \textbf{Uncertainty}  \\
\noalign{\smallskip}\hline\noalign{\smallskip}
		$b_0$ & $-3$ & $\phantom{-}1$\\
		$b_1$ & $\phantom{-}26$ & $\phantom{-}4$\\
		$b_2$ & $\phantom{-}0.35$ & $\phantom{-}0.02$\\
		$b_3$ & $\phantom{-}12.7$ & $\phantom{-}0.1$\\
		$b_4$ & $-0.0039$ & $\phantom{-}0.0008$\\
\noalign{\smallskip}\hline
\end{tabular}
\end{table}
\section{Measurement of the Energy Spectrum}
\label{s:Spectrum}

To build the energy spectrum from the reconstructed energy distribution, we need to correct the raw spectrum, obtained as $J^\text{raw}_i=N_i/(\mathcal{E}\Delta E_i)$, for the bin-to-bin migrations of events due to the finite accuracy with which the energies are assigned. 
The energy bins are chosen to be regularly sized in decimal logarithm, $\Delta\lg E_i=0.1$, commensurate with the energy resolution. 
The level of migration is driven by the resolution function, the detection efficiency in the energy range just below the threshold energy, and the steepness of the spectrum.
To correct for these effects, we use the bin-by-bin correction approach presented in~\cite{Aab:2020gxe}. 
It consists of folding the detector effects into a \textit{proposed} spectrum function, $J(E,\mathbf{k})$, with free parameters, $\mathbf{k}$, such that the result describes the set of the observed number of events $N_i$.
The set of expectations, $\nu_i$, is obtained as $\nu_i(\mathbf{k})=\sum_j R_{ij}\mu_j(\mathbf{k})$,
where the $R_{ij}$ coefficients (reported in a matrix format in the Supplementary material) describe the bin-to-bin migrations, and where $\mu_j$ are the expectations in the case of an ideal detector obtained by integrating the proposed spectrum over $E_j$ and $E_j+\Delta E_j$ scaled by $\mathcal{E}$. 
The optimal set of free parameters, $\hat{\mathbf{k}}$, is inferred by minimizing a log-likelihood function built from the Poisson probabilities to observe $N_i$ events when $\nu_i(\hat{\mathbf{k}})$ are expected. 

\begin{figure}[t]
\centering
\includegraphics[width=0.99\columnwidth]{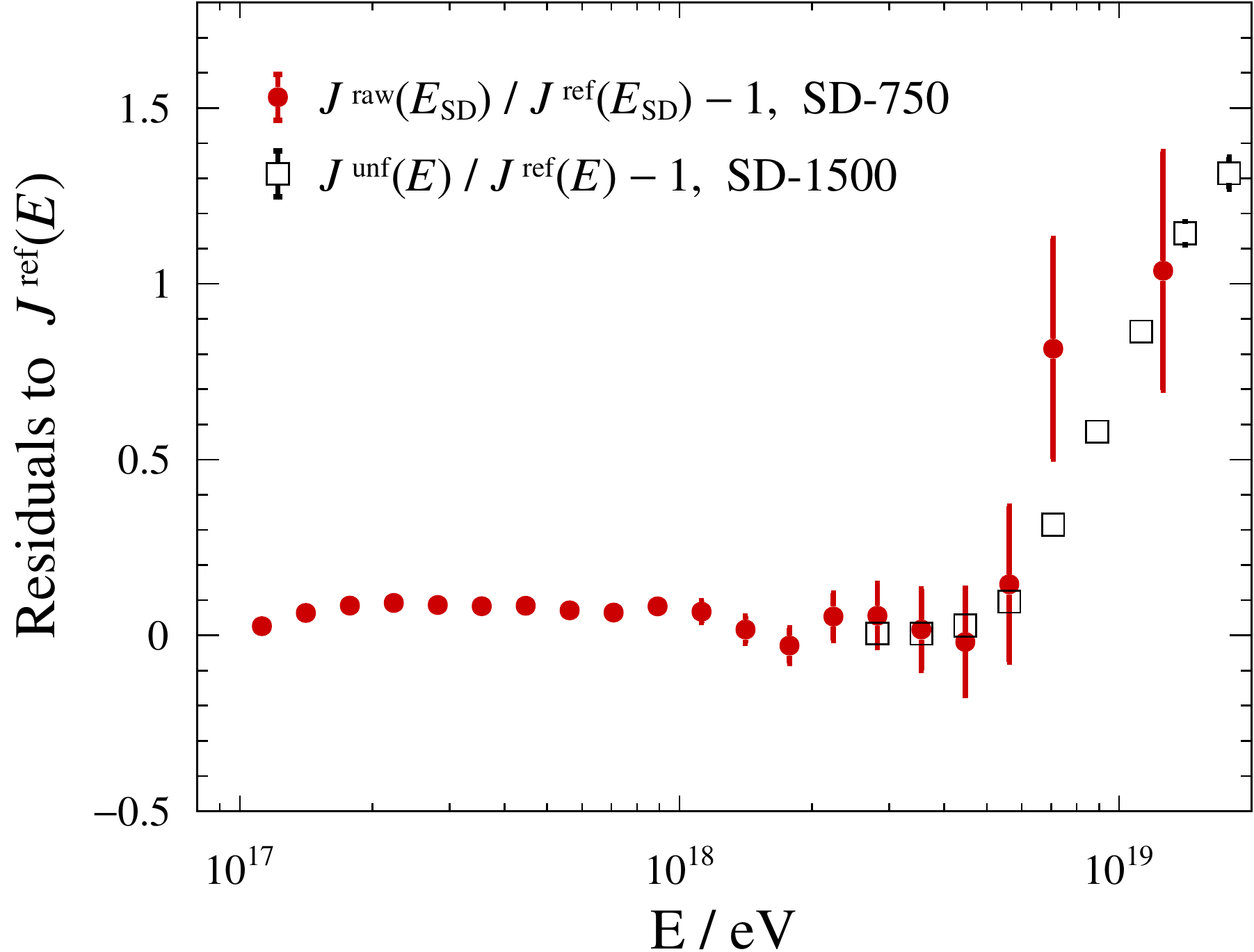}
\caption{Residuals of the SD-750 raw spectrum with respect to the power-law function $J^\text{ref}(E)$. Data points from the SD-1500 spectrum measurement are superimposed.}
\label{f:Residuals}
\end{figure}

To choose the proposed function, we plot in \cref{f:Residuals} the residuals (red dots) of the SD-750 raw spectrum with respect to a reference function, $J^\text{ref}(E)$, that fits the SD-1500 spectrum below the ankle energy down to the SD-1500 threshold energy, $10^{18.4}$\,eV.
A re-binning was applied at and above $10^{19}$\,eV to avoid too large statistical fluctuations.

The reference function in this energy range, as reported in~\cite{Aab:2020gxe}, is 
\begin{equation}
J^\text{ref}(E)=J_0^\text{ref}\left(\frac{E}{10^{18.5}\,\text{eV}}\right)^{-\gamma_1^\text{ref}}, 
\label{eqn:Jref}
\end{equation}
with $J_0^\text{ref}=1.315{\times}10^{-18}$\,km$^{-2}$\,yr$^{-1}$\,sr$^{-1}$\,eV$^{-1}$ and $\gamma_1^\text{ref}=3.29$.     
The residuals of the SD-1500 unfolded spectrum with respect to $J^\text{ref}(E)$ are also shown as open squares in \cref{f:Residuals}. 
The sharp transition at ${\simeq}10^{18.7}$\,eV to a different power law corresponds to the spectral feature known as the \textit{ankle}.
Such a transition is also observed, with much lower sensitivity, using data from the SD-750 array.
Below ${\simeq}10^{18.7}$\,eV and down to ${\simeq}10^{17.4}$\,eV, one can see a shift of the raw SD-750 spectrum compared to $J^\text{ref}(E)$.
This is expected from a combination of primarily the resolution effects to be unfolded and of a possible mismatch, within the energy-dependent budget of \emph{uncorrelated} uncertainties, of the SD-1500 and SD-750 $E_\text{SD}$ energy scales.
Below ${\simeq}10^{17.4}$\,eV, a slight roll-off begins. 
Overall, these residuals are suggestive of a power-law function to describe the data leading up to the ankle energy where the spectrum hardens, with a gradually changing spectral index over the lowest energies studied.
Consequently, the proposed function is chosen as three power laws with transitions occurring over adjustable energy ranges, 
\begin{equation}
J(E,\mathbf{k}) = J_0 \left(\frac{E}{10^{17}\,\text{eV}}\right)^{-\gamma_0} \prod_{i=0}^1\left[1+\left(\frac{E}{E_{ij}}\right)^{\frac{1}{\omega_{ij}}}\right]^{(\gamma_i-\gamma_j)\omega_{ij}},
\label{eqn:J}
\end{equation}
with $j=i+1$.
The normalization factor $J_0$, the three spectral indices $\gamma_i$, and the transition parameter $\omega_{01}$ constitute the free parameters in $\mathbf{k}$. 
The transition parameter $\omega_{12}$, constrained with much more sensitivity using data from the SD-1500, is fixed at $\omega_{12}=0.05$~\cite{Aab:2020gxe}.

\begin{figure}[t]
\centering
\includegraphics[width=0.99\columnwidth]{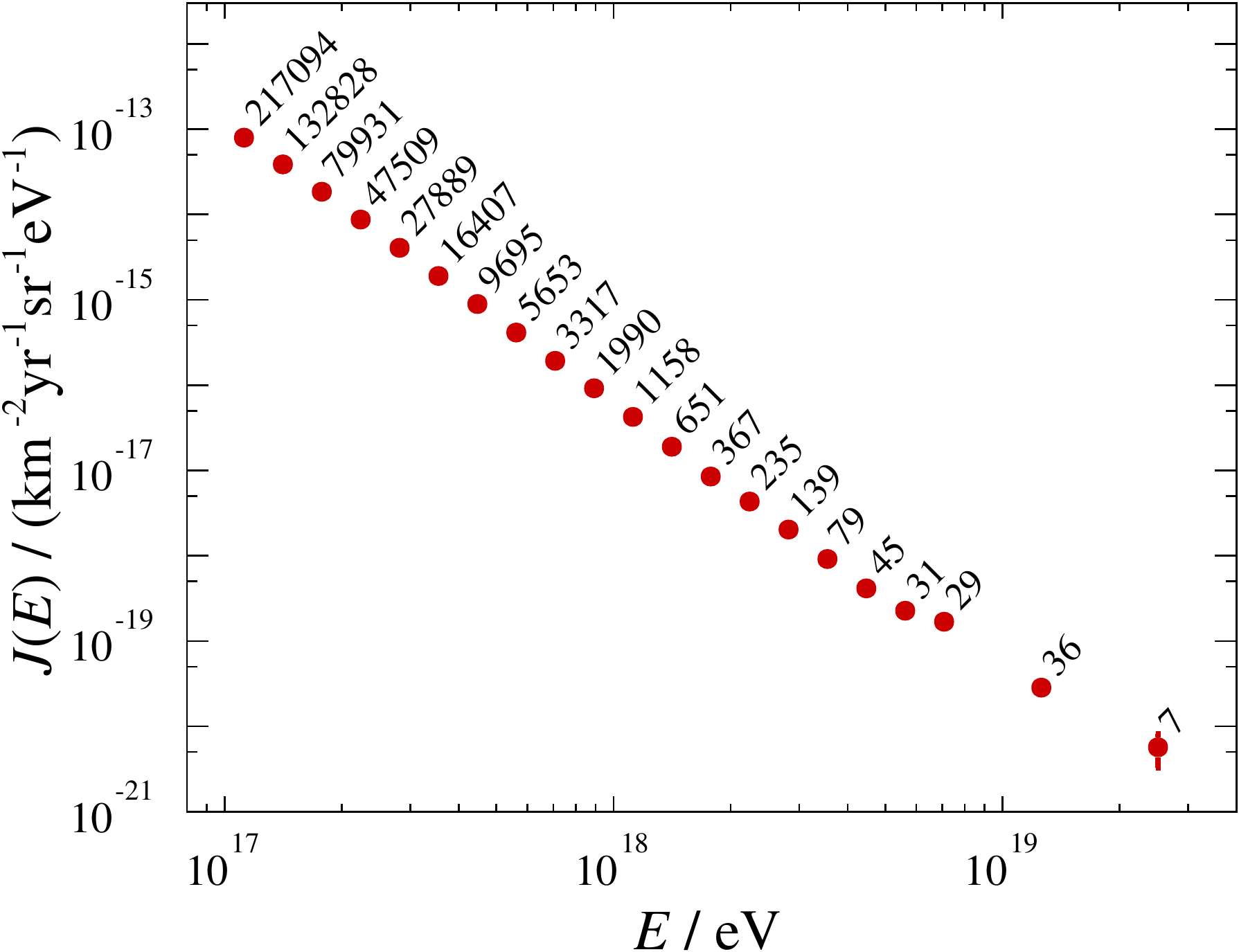}
\caption{Unfolded energy spectrum derived using data from the SD-750 array.}
\label{f:UnfSpectrum}
\end{figure}

\begin{table}
\caption{Best-fit values of the spectral parameters (\cref{eqn:J}). The parameter $\omega_{12}$ is fixed to the value constrained in~\cite{Aab:2020gxe}. Note that the parameters $\gamma_0$ and $E_{01}$ correspond to features below the measured energy region and are treated only as aspects of the unfolding fixed to their best-fit values to infer the uncertainties of the measured spectral parameters.}
\label{tab:spectral_parameters}       
\begin{tabular}{lll}
\hline\noalign{\smallskip}
\textbf{Parameter} & \textbf{Value $\pm \sigma_\text{stat} \pm \sigma_\text{syst}$}  \\
\noalign{\smallskip}\hline\noalign{\smallskip}
		$J_0/($km$^{2}$\,yr\,sr\,eV) & $\left(1.09 \pm 0.04 \pm 0.28 \right){\times} 10^{-13}$  \\
		$\omega_{01}$ & $\phantom{(}0.49\pm0.07 \pm 0.34 $ \\
		$\gamma_1$ &  $\phantom{(}3.34\pm 0.02 \pm 0.09 $ \\
		$E_{12}/$eV & $\left(3.9\pm 0.8 \pm  1.1 \right){\times} 10^{18}$  \\
		$\gamma_2$ & $\phantom{(}2.6\pm 0.2 \pm 0.1 $ \\
		&\\[-0.8em]
		\hline\noalign{\smallskip}
		$\gamma_0$ & $2.64$  -- fixed \\
		$E_{01}/$eV &  $1.24{\times}10^{17}$ -- fixed \\
		$\omega_{12}$ & 0.05 -- fixed \\
\noalign{\smallskip}\hline
\end{tabular}
\end{table}

Combining all the ingredients at our disposal, we obtain the final estimate of the spectrum, $J_i$, unfolded for the effects of the response of the detector and shown in \cref{f:UnfSpectrum}.
It is obtained as
\begin{equation}
    \label{eqn:Junf}
    J_i=\frac{\mu_i}{\nu_i}J^\text{raw}_i = c_i\,J^\text{raw}_,
\end{equation}
where the $\mu_i$ and $\nu_i$ coefficients are estimated using the best-fit parameters $\hat{\mathbf{k}}$. 
Their ratios define the bin-by-bin corrections used to produce the unfolded spectrum.
The correction applied extends from 0.84 at $10^{17}$\,eV to 0.99 around the ankle (see \ref{app::spectrumdata}).
The best-fit spectral parameters are reported in \cref{tab:spectral_parameters}, while the statistical correlations between the parameters are detailed in \ref{app::spectrumdata} (\cref{tab:rho_param750stat}).
The goodness-of-fit of the forward-folding procedure is attested by the deviance of $15.9$, which, if considered to follow the C statistics~\cite{Bonamente:2019efn}, can be compared\footnote{Note that the $p$-value for a proposed function which does not include a transition from $\gamma_0$ to $\gamma_1$ can be rejected with more than $20\sigma$ confidence.} to the expectation of $16.2\pm 5.6$ to yield a $p$-value of $0.50$.

\begin{figure}[t]
\centering
\includegraphics[width=0.99\columnwidth]{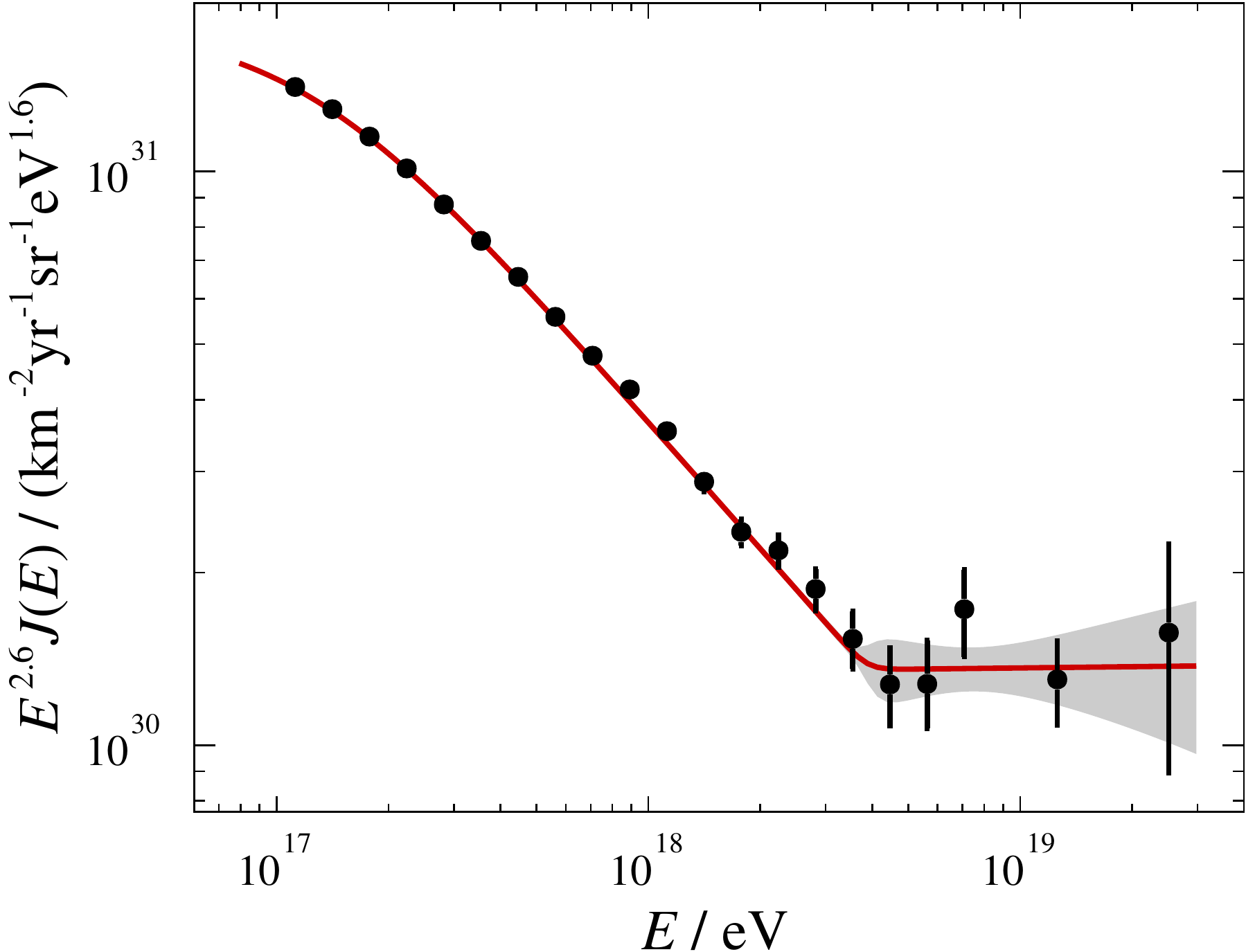}
\caption{Unfolded energy spectrum of the SD-750, scaled by $E^{2.6}$.}
\label{f:UnfScaledSpectrum}
\end{figure}

The fitting function is shown in \cref{f:UnfScaledSpectrum}, superimposed to the spectrum scaled by $E^{2.6}$, allowing one to better appreciate its characteristics, from the turn-over at around $10^{17}$\,eV up to a few $10^{19}$\,eV, thus including the ankle. 
The turn-over is observed with a very large exposure, unprecedented at such energies. 
However, as indicated by the magnitude of the transition parameter, $\omega_{01}\simeq0.49$, the change of the spectral index occurs over an extended $\Delta\lg E\simeq 0.5$ energy range, so that the spectral index $\gamma_0$ cannot be observed but only indirectly inferred. 
Also, the value of the energy break, $E_{01}\simeq 1.24{\times}10^{17}$\,eV, turns out to be close to the threshold energy. 
These two facts thus imply that, while a spectral break is found beyond any doubt, it cannot wholly be characterised, as only the higher energy portion is actually observed. 
Consequently, the fit values describing $E_{01}$ and $\gamma_0$ are not to be considered as true measurements but as necessary parameters in the fit function, the statistical resolutions of which are on the order of 35\%. 
Once we infer their best-fit values, we use these values as ``external parameters'' to estimate the uncertainties of the other spectral parameters.
This procedure gives rise to an increase of the systematic uncertainties, but is necessary as $E_{01}$ and $\gamma_0$ are not directly observed.
Beyond the smooth turn-over around $E_{01}$, the intensity can be described by a power-law shape as $J(E)\propto E^{-\gamma_1}$, up to $E_{12} = \left(3.9\pm 0.8\right){\times}10^{18}$\,eV, the ankle energy, the value of which is within 1.4$\sigma$ of that found with the much larger exposure of the SD-1500 measurement of the spectrum, namely $(5.0\pm0.1){\times}10^{18}$\,eV. 
Also the value of $\gamma_1 = 3.34\pm 0.02$ is within 1.8$\sigma$ of that obtained with the SD-1500 between $10^{18.4}$ and $10^{18.7}$\,eV ($3.29 \pm 0.02$).

\begin{figure}[t]
\centering
\includegraphics[width=0.99\columnwidth]{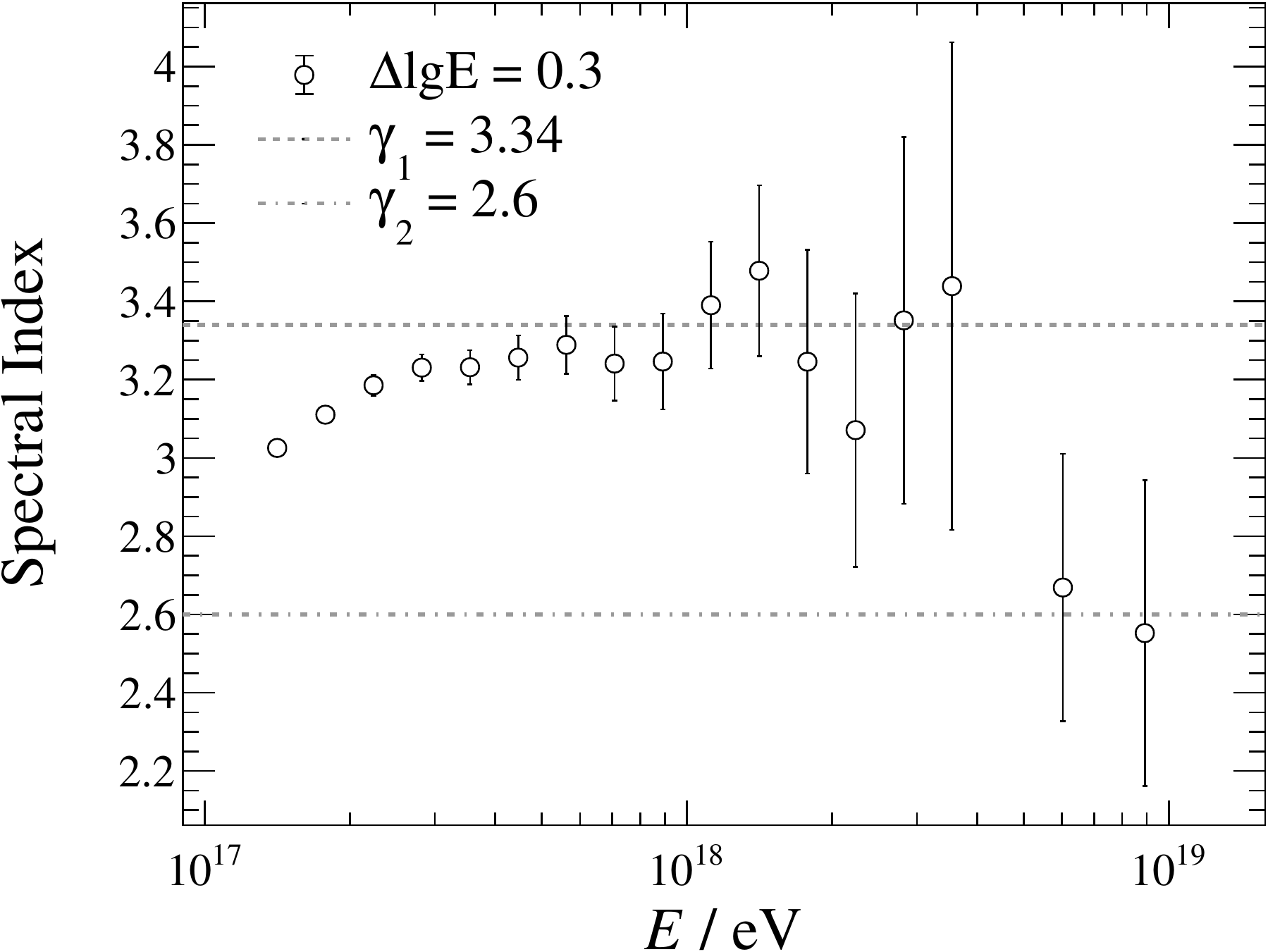}
\caption{Evolution of the spectral index with energy. The measured spectral points were fit to power laws within a sliding window of $\Delta\lg E = 0.3$. The values of $\gamma_1$ and $\gamma_2$ are represented by the dashed and dash-dotted lines, for reference.}
\label{f:SpectralIndex}
\end{figure}

The characteristics of the measured spectrum can also be studied by looking at the evolution of the spectral index as a function of energy, $\gamma(E)$.
Rather than relying on the empirically chosen unfolding function, this slope parameter can be directly fit using the values calculated in $J(E)$.
Power-law fits were performed for a sliding window of width $\Delta\lg E = 0.3$. The resulting estimations of the so obtained spectral indexes are shown in \cref{f:SpectralIndex}.
The values of the spectral index fits present a consistent picture of the evolution.
Beginning at the lowest energies shown, $\gamma(E)$ increases first quite rapidly, finally approaching a value of 3.3 leading up to the ankle asymptotically.
Unsurprisingly, this is the value found for $\gamma_1$ in the unfolding of both the SD-750 and SD-1500 spectra~\cite{Aab:2020gxe}.

The systematic uncertainties that affect the measurement of the spectrum are dominated by the overall uncertainty of the energy scale, detailed in~\cite{energy_scale}, and is, itself, dominated by the absolute calibration of the fluorescence telescopes (10\%). 
The total uncertainty in the energy scale is $\sigma_E / E = 14$\%.
Once propagated, the steepness of the spectrum as a function of energy amplifies this uncertainty, roughly as $\sigma_{J}/J = (\gamma_1 - 1)\sigma_E / E$, resulting in a total flux uncertainty of $\sigma_{J}/J \simeq 35$\%.
However, for a more exact calculation of the uncertainty, the energies of the individual events were shifted by $\pm14$\% and the unfolding procedure was repeated.
%
\begin{figure}[t]
\centering
\includegraphics[width=0.99\columnwidth]{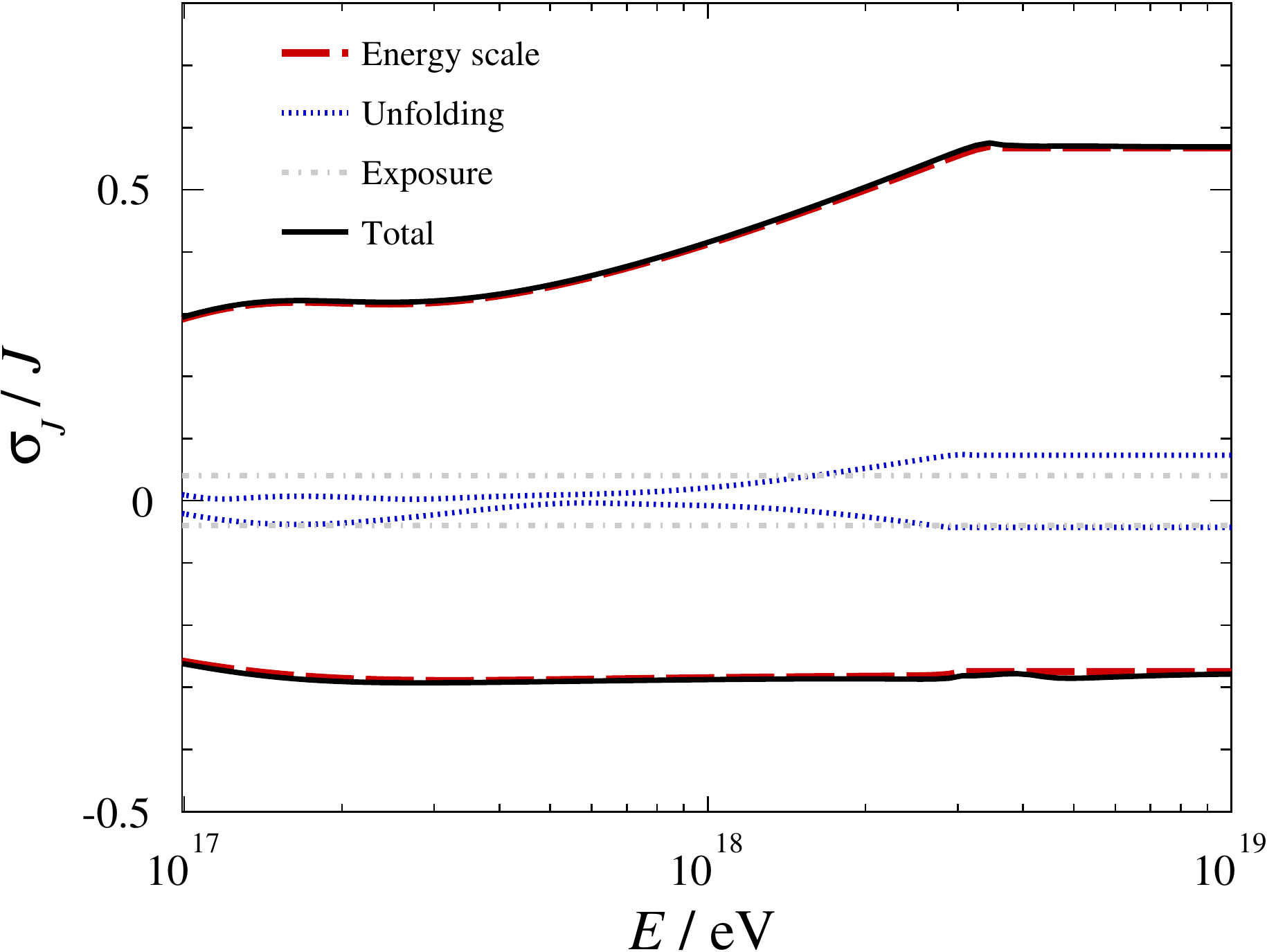}
\caption{Systematic uncertainties in the flux measurement as a function of energy. The main contributions are shown separately.}
\label{f:Systematics}
\end{figure}
The result is shown as dashed red lines in \cref{f:Systematics}.

Beyond that of the energy scale, the additional uncertainties are subdominant but are important to understand as they have energy dependence and some are uncorrelated with other flux measurements made at the Observatory.
Such knowledge is particularly important for the combination of the two SD spectra presented later in~\cref{s:combination}.
The most relevant of these energy-dependent uncertainties is associated with the procedure of the forward-folding itself.
The uncertainties in the resolution function and in the detection efficiency all contribute a component to the overall unfolding uncertainty.
The forward-folding process was hence repeated by shifting,
within the statistical uncertainties, the parameterizations of the energy resolution (\cref{eq:EnergyResolution}) and efficiency parameterization, and by bracketing the bias with the pure proton/iron mass primaries below full efficiency. 
The impact of the resolution uncertainties on the unfolding procedure is the larger, in particular at the highest energies.
On the other hand, the energy bias and reduced efficiency below $10^{17}$\,eV only impacts the first few bins.
These various components are summed in quadrature and are shown by the dotted blue line in \cref{f:Systematics}.
These influences are clearly seen to impact the spectrum by ${<}4\%$. 

The last significant uncertainty in the flux is related to the calculation of the geometric exposure of the array.
This quantity has been previously studied and is 4\% for the SD-750 which directly translates to a 4\% energy-independent shift in the flux~\cite{ThePierreAuger:2015rma}.

The resulting systematic uncertainties of the spectral parameters are given in \cref{tab:spectral_parameters}.
For completeness, beyond the summary information provided by the spectrum parameterization, the correlation matrix of the energy spectrum is given in the Supplementary material.
It is obtained by repeating the analysis on a large number of data sets, sampling randomly the systematic uncertainties listed above.

\section{The Combined SD-750 and SD-1500 Energy Spectrum}
\label{s:combination}

\begin{figure}[t]
\centering
\includegraphics[width=0.99\columnwidth]{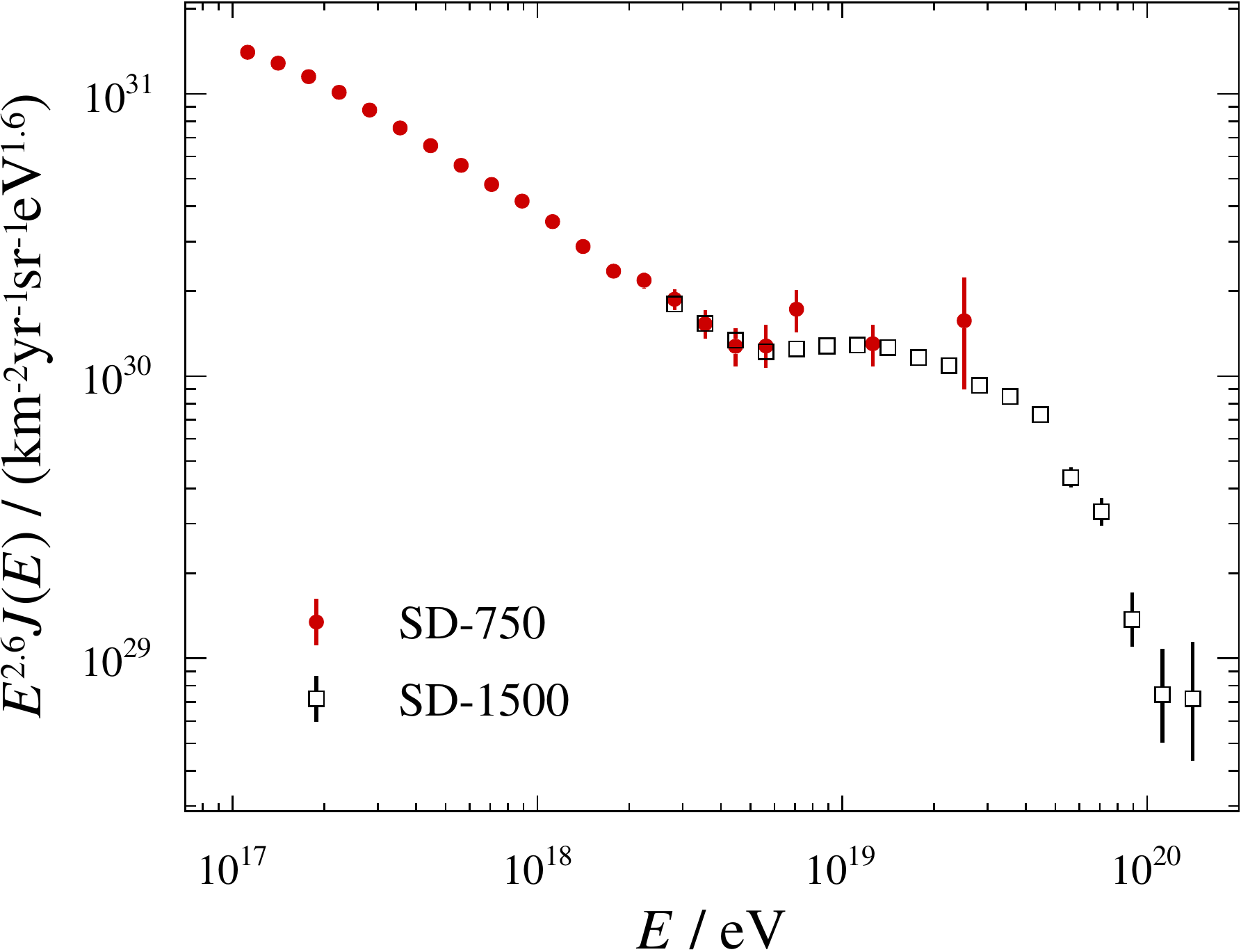}
\caption{Superimposed SD spectra to be combined scaled by $E^{2.6}$, the SD-750 (red circles) and the SD-1500 (black squares).}
\label{f:UnfSpectra}
\end{figure}

The spectrum obtained in \cref{s:Spectrum} extends down to $10^{17}$\,eV and at the high-energy end overlaps with the one recently reported in~\cite{Aab:2020gxe} using the SD-1500 array. 
The two spectra are superimposed in \cref{f:UnfSpectra}.
Beyond the overall consistency observed between the two measurements, a combination of them is desirable to gather the information in a single energy spectrum above $10^{17}$\,eV obtained with data from both the SD-750 and the SD-1500 of the Pierre Auger Observatory.
We present below such a combination considering adjustable re-scaling factors in exposures, $\delta\mathcal{E}$, and $E_\text{SD}$ energy scales, $\delta E_\text{SD}$, within uncorrelated uncertainties.

%
\begin{figure*}[t]
\centering
\includegraphics[width=1.6\columnwidth]{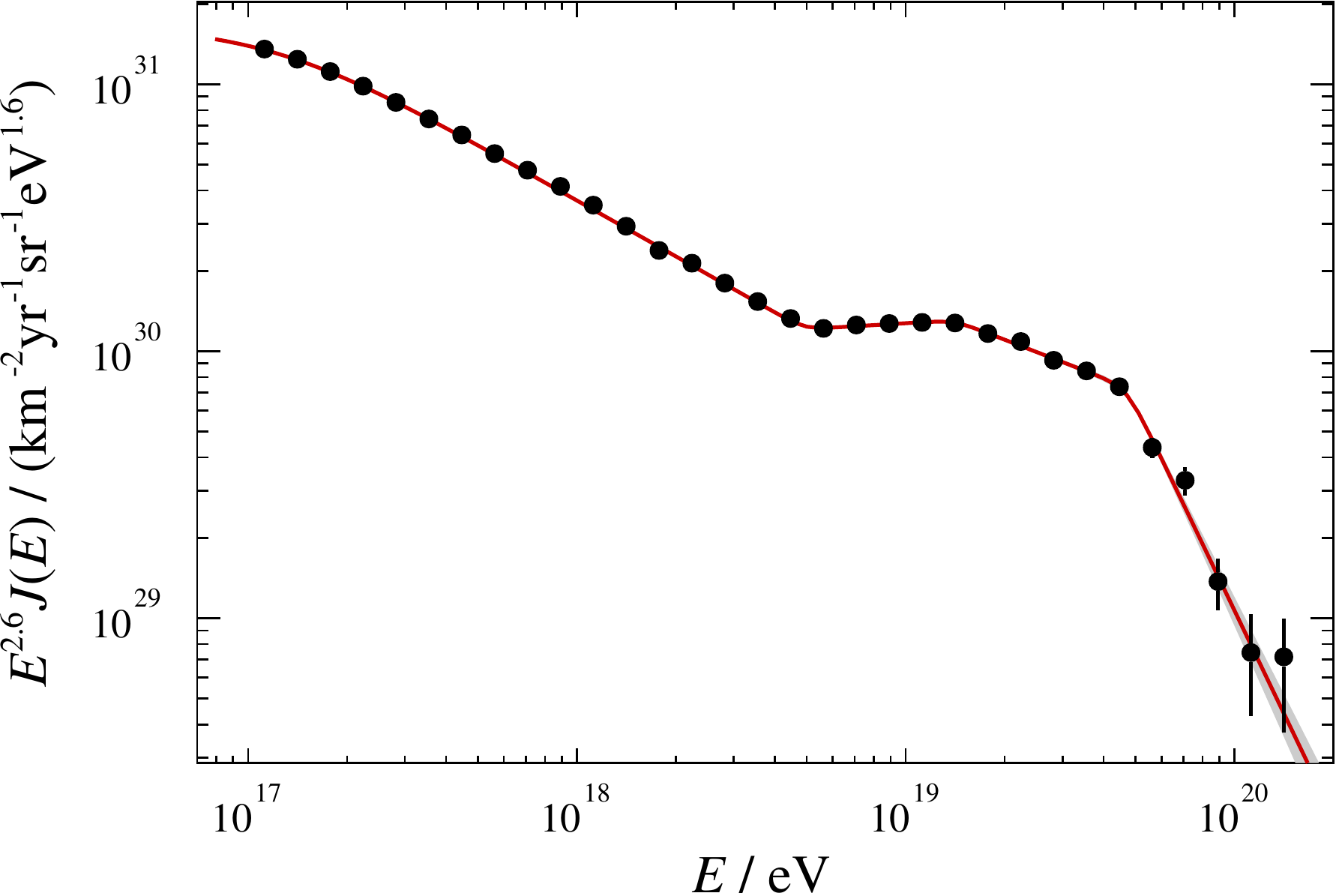}
\caption{SD energy spectrum after combining the individual measurements by the SD-750 and the SD-1500 scaled by $E^{2.6}$. The fit using the proposed function (\cref{eqn:Jfunc}) is overlaid in red along with the one sigma error band in gray.} 
\label{f:CombinedSpectra}
\end{figure*}

The combination is carried out using the same bin-by-bin correction approach as in \cref{s:Spectrum}. 
The joint likelihood function, $\mathcal{L}(\mathbf{s},\delta\mathcal{E},\delta E_\text{SD})$, is built from the product of the individual Poissonian likelihoods pertaining to the two SD measurements, $\mathcal{L}_{750}$ and $\mathcal{L}_{1500}$. 
These two individual likelihoods share the same proposed function,
\begin{equation}
J(E,\mathbf{s}) = J_0 \left(\frac{E}{E_0}\right)^{-\gamma_0} \frac{\prod_{i=0}^3\left[1+\left(\frac{E}{E_{ij}}\right)^{\frac{1}{\omega_{ij}}}\right]^{(\gamma_i-\gamma_j)\omega_{ij}}}{\prod_{i=0}^3\left[1+\left(\frac{E_0}{E_{ij}}\right)^{\frac{1}{\omega_{ij}}}\right]^{(\gamma_i-\gamma_j)\omega_{ij}}},
\label{eqn:Jfunc}
\end{equation}
with $j=i+1$ and $E_0 = 10^{18.5}$\,eV.
As in~\cite{Aab:2020gxe}, the transition parameters $\omega_{12}$, $\omega_{23}$ and $\omega_{34}$ are fixed to 0.05.
In this way, the same parameters $\mathbf{s}$ are used during the minimisation process to calculate the set of expectations $\nu_i(\mathbf{s},\delta\mathcal{E},\delta E_\text{SD})$ of the two arrays. 
For each array, a change of the associated exposure $\mathcal{E}\to\mathcal{E}+\delta\mathcal{E}$ impacts the $\nu_i$ coefficients accordingly, while a change in energy scale $E_\text{SD}\to E_\text{SD}+\delta E_\text{SD}$ impacts as well the observed number of events in each bin.  
Additional likelihood factors, $\mathcal{L}_{\delta\mathcal{E}}$ and $\mathcal{L}_{\delta E_\text{SD}}$, are thus required to control the changes of the exposure and of the energy-scale within their uncorrelated uncertainties. 
The likelihood factors described below account for $\delta\mathcal{E}$ and $\delta E_\text{SD}$ changes associated with the SD-750 only.
We have checked that allowing additional free parameters, such as the $\delta\mathcal{E}$ corresponding to the SD-1500, does not improve the deviance of the best fit by more than one unit, and thus their introduction is not supported by the data. 

Both likelihood factors are described by Gaussian distributions with a spread given by the uncertainty pertaining to the exposure and to the energy-scale.
The joint likelihood function reads then as
\begin{equation}
    \mathcal{L}(\mathbf{s},\delta\mathcal{E},\delta E_\text{SD})=\mathcal{L}_{750}\times \mathcal{L}_{1500}\times \mathcal{L}_{\delta\mathcal{E}}\times \mathcal{L}_{\delta E_\text{SD}}. 
\end{equation}
The allowed change of exposure, $\delta\mathcal{E}$, is guided by the systematic uncertainties in the SD-750 exposure, $\sigma_\mathcal{E}/\mathcal{E}=4\%$.
Hence, the constraining term for any change in the SD-750 exposure reads, dropping constant terms, as
\begin{equation}
    -2\ln \mathcal{L}_{\delta\mathcal{E}}(\delta\mathcal{E}) = \left(\frac{\delta\mathcal{E}}{\sigma_\mathcal{E}}\right)^2.
\end{equation}
Likewise, uncertainties in $A$ and $B$, $\delta A$ and $\delta B$, translate into uncertainties in the SD-750 energy scale. 
Statistical contributions stem from the energy calibration of $S_{35}$, which are by essence uncorrelated to those of the SD-1500. 
Other uncorrelated contributions of the systematic uncertainties from the FD energy scales propagated to the SD-1500 and SD-750 could enter into play. 
The magnitude of such systematics, $\sigma_{\mathrm{syst}}$, is difficult to quantify. 
By testing several values for $\sigma_{\mathrm{syst}}$, we have checked, however, that such contributions have a negligible impact on the combined spectrum.
Hence, the constraining term for any change in energy scale can be considered to stem from statistical uncertainties only and reads as
\begin{eqnarray}
    -2\ln \mathcal{L}_{E_{\mathrm{SD}}}(\delta A,\delta B) &=& [\sigma^{-1}]_{AA}(\delta A)^2+[\sigma^{-1}]_{BB}(\delta B)^2 \nonumber \\
    &\quad&+\,2[\sigma^{-1}]_{AB}(\delta A)(\delta B),
\end{eqnarray}
where the notation $[\sigma]_{ij}$ stands for the coefficients of the variance-covariance matrix of the $A$ and $B$ best-fit estimates and $[\sigma^{-1}]$ is the inverse of this matrix.

\begin{table}
\caption{Best-fit values of the combined spectral parameters (\cref{eqn:Jfunc}). The parameter $\omega_{12}$, $\omega_{23}$ and $\omega_{34}$ are fixed to the value constrained in~\cite{Aab:2020gxe}. Note that the parameters $\gamma_0$ and $E_{01}$ correspond to features below the measured energy region and should be treated only as aspects of the combination.}
\label{tab:combined_parameters}       
\begin{tabular}{lll}
\hline\noalign{\smallskip}
\textbf{Parameter} & \textbf{Value $\pm \sigma_\text{stat} \pm \sigma_\text{syst}$}  \\
\noalign{\smallskip}\hline\noalign{\smallskip}
		$J_0$\,/\,(km$^{2}$\,yr\,sr\,eV) & $\left(1.309 \pm 0.003 \pm 0.400 \right){\times} 10^{-18}$  \\
		$\omega_{01}$ & $\phantom{(}0.43\pm 0.04 ~\pm 0.34 $ \\
		$\gamma_1$ &  $\phantom{(}3.298\pm 0.005 \pm 0.10 $ \\
		$E_{12} / \text{eV}$ & $\left(4.9\pm 0.1 \pm 0.8 \right){\times} 10^{18}$  \\
		$\gamma_2$ & $\phantom{(}2.52\pm 0.03 \pm 0.05 $ \\
		$E_{23} / \text{eV}$ & $\left(1.4\pm 0.1 \pm 0.2 \right){\times} 10^{19}$ \\
		$\gamma_3$ & $\phantom{(}3.08\pm 0.05 \pm 0.10 $ \\
		$E_{34} / \text{eV}$ & $\left(4.7\pm 0.3 \pm 0.6 \right){\times} 10^{19}$ \\
		$\gamma_4$ & $\phantom{(}5.2\pm 0.2 \pm 0.1  $ \\
		\hline\noalign{\smallskip}
		$\gamma_0$ & $2.64 $ -- fixed  \\
		$E_{01} / \text{eV}$ &  $1.24{\times} 10^{17} $ -- fixed \\
		$\omega_{12}$ & 0.05 -- fixed \\
		$\omega_{23}$ & 0.05 -- fixed\\
		$\omega_{34}$ & 0.05 -- fixed \\
\noalign{\smallskip}\hline
\end{tabular}
\end{table}

The outcome of the forward-folding fit is the set of parameters $\mathbf{s}$, $\delta\mathcal{E}$, $\delta A$ and $\delta B$ that allow us to calculate the expectation values $\mu_i$ and $\nu_i$, and thus the correction factors $c_i$, for both arrays separately.
The resulting combined spectrum, obtained as
\begin{equation}
    \label{eqn:Jcomb}
    J^\text{comb}_i=\frac{c_{i,750}\,N_{i,750}+c_{i,1500}\,N_{i,1500}}{\mathcal{E}_i^\text{eff}\,\Delta E_i},
\end{equation}
is shown in \cref{f:CombinedSpectra}. 
Here, the observed number of events $N_i^{750}$ in each bin is calculated at the re-scaled energies, while the effective exposure, $\mathcal{E}_i^\text{eff}$, is the shifted one of the SD-750 in the energy range where $N_{i,1500}=0$, the one of the SD-1500 in the energy range where $N_{i,750}=0$, and the sum $\mathcal{E}_{750}+\delta\mathcal{E}+\mathcal{E}_{1500}$ in the overlapping energy range.
The set of spectral parameters are collected in \cref{tab:combined_parameters}, while the corresponding correlation matrix is reported in~\ref{app::spectrumdata} (\cref{tab:rho_paramSDstat}) for $\delta\mathcal{E}$, $\delta A$ and $\delta B$ fixed to their best-fit values.
The change in exposure is $\delta\mathcal{E}/\mathcal{E}=+1.4\%$, while the one in energy scale follows from $\delta A/A=-2.5\%$ and $\delta B/B=+0.8\%$.
The goodness-of-fit is evidenced by a deviance of 37.2 for an expected value of $32\pm8$. We also note that the parameters describing the spectral shape are in agreement with those of the two individual spectra from the SD arrays.

The impact of the systematic uncertainties, dominated by those in the energy scale, on the spectral parameters are reported in \cref{tab:combined_parameters}.
For completeness, beyond the summary information provided by the spectrum parameterization, the correlation matrix of the energy spectrum itself is also given in the Supplementary material. 
\section{Discussion}
\label{s:conclusion}

\begin{figure*}[t]
\centering
\includegraphics[width=1.6\columnwidth]{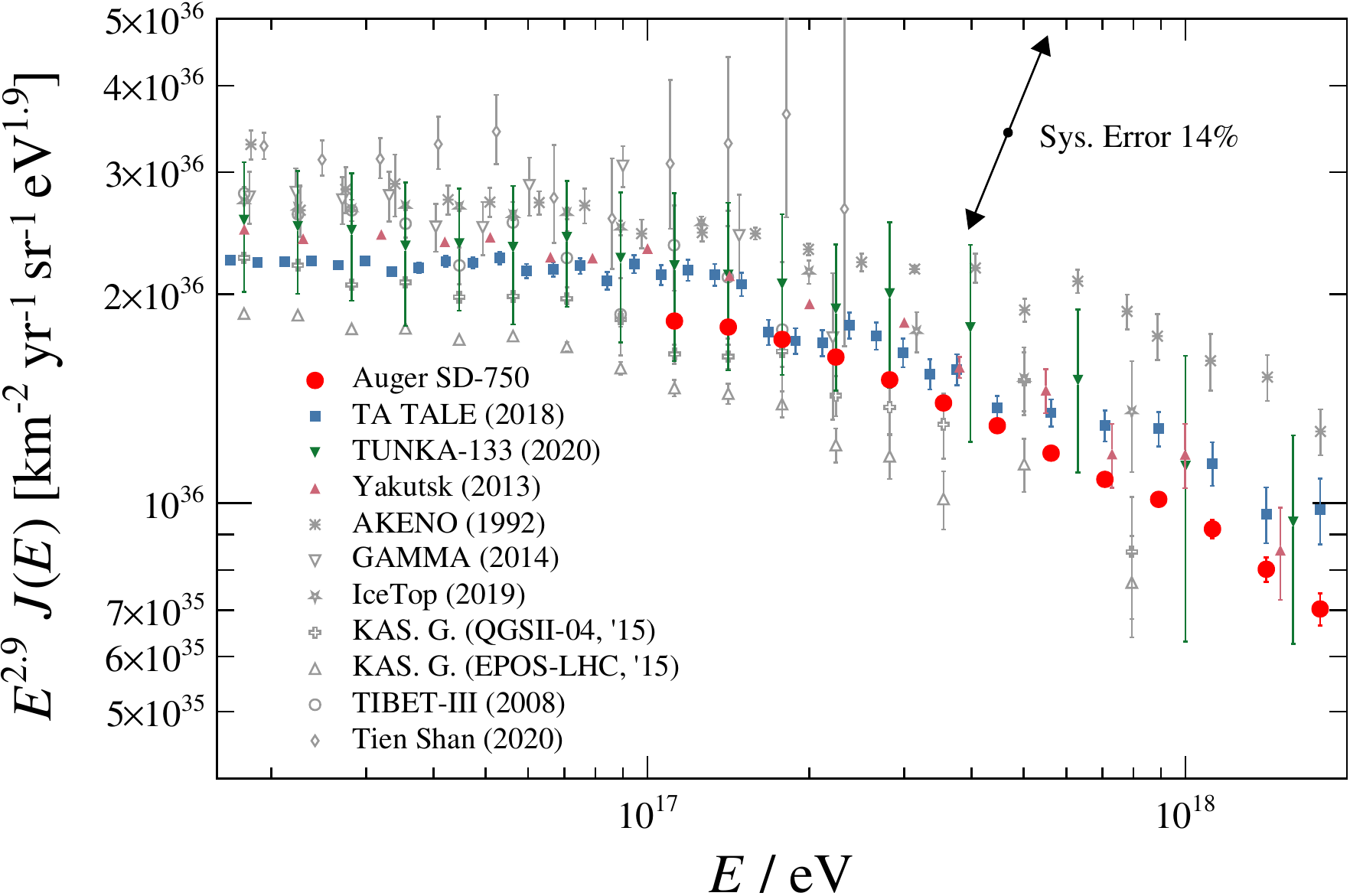}
\caption{SD-750 spectrum (solid red circles) near the second knee along with the measurements from Akeno~\cite{akeno1992spec}, GAMMA~\cite{gamma2007spec}, IceTop~\cite{IceCube:2019hmk}, KASCADE-Grande~\cite{Bertaina:2015fnz}, TALE~\cite{Abbasi:2018xsn}, Tien Shan~\cite{gudkova2020results}, Tibet-III~\cite{amenomori2008all}, Tunka-133~\cite{Budnev:2020oad}, Yakutsk~\cite{knurenko2013cosmic}. The experiments that set their energy scale using calorimetric observations are indicated by solid colored markers while those with an energy scale based entirely on simulations are shown by gray markers.}
\label{f:AllAtSecondKnee}
\end{figure*}

We have presented here a measurement of the CR spectrum in the energy range between the second knee and the ankle, which is covered with high statistics by the SD-750, including 560,000 events with zenith angles up to $40^\circ$ and energies above $10^{17}$\,eV.
The measurement includes a total exposure of $105$\,km$^2$\,sr\,yr and an energy scale set by calorimetric observations from the FD telescopes.
We note a significant change in the spectral index and with a width that is much broader than that of the ankle feature.

%
%

Such a change has been observed by a number of other experiments, and via various detection methods.
Most notably, the nature of this feature was linked to a softening of the heavy-mass primaries beginning at $10^{16.9}$\,eV by the KASCADE-Grande experiment, leading to the moniker \emph{iron knee}~\cite{Apel:2012tda}.
Additional analyses by the Tunka-133~\cite{gress1999study} and IceCube~\cite{IceCube:2019hmk} collaborations have given further evidence that high-mass particles are dominant near $10^{17}$\,eV and thus that it is their decline that largely defines the shape of the all-particle spectrum.
The hypothesis is also supported by a preliminary study of the distributions of the depths of the shower maximum, $X_\text{max}$, measured at the Auger Observatory~\cite{Aab:2014aea,Bellido:2017cgf}. These have been parametrized according to the hadronic models EPOS-LHC~\cite{Pierog:2013ria},
QGSJetII-04~\cite{ostapchenko2013qgsjet} and Sibyll2.3~\cite{riehn2017hadronic}. From these parametrizations, the evolution over energy of the fractions of different mass groups, from protons to Fe-nuclei, has been derived. From all three models, a fall-off of the Fe component above $10^{17}$\,eV is inferred. The consistency of all these observations strongly supports a scenario of Galactic CRs characterised by a rigidity-dependent maximum acceleration energy for particles with charge $Z$, namely $E_\text{max}(Z)\simeq ZE_\text{max}^\text{proton}$, to explain the knee structures. 

The measurements of the all-particle flux from various experiments~\cite{akeno1992spec, gamma2007spec, IceCube:2019hmk, Bertaina:2015fnz, Abbasi:2018xsn, gudkova2020results, amenomori2008all, Budnev:2020oad, knurenko2013cosmic} in the energy region surrounding the second knee are shown in \cref{f:AllAtSecondKnee}.
Experiments which set their energy scale using calorimetric measurements are plotted using colored markers (Auger SD-750, TA TALE, TUNKA-133, Yakutsk) while the measurements shown in gray markers represent MC-based energy assignments.
The spread between various experiments is statistically significant.
However, all these measurements are consistent with the SD-750 spectrum within the 14\% energy scale systematic uncertainty.
Understanding the nature of the off-sets in the energy scales is beyond the scope of this paper.
However, we note that the TALE spectrum agrees rather well with the SD-750 spectrum, offset by 5 to 6\% in energy.
The agreement is notable given that at-and-above the ankle, an energy scale off-set of around 11\% is required to bring the spectral measurements with SD-1500 of the Auger Observatory and the SD of the Telescope Array into agreement~\cite{Deligny:2019SC}.

Additionally, we have presented a robust method to combine energy spectra.
Using the result from the SD-750 and a previously reported measurement using the SD-1500, a unified SD spectrum was calculated by combining the respective observed fluxes, energy resolutions, and exposures.
The result has partial coverage of the second knee and full coverage of the ankle, an additional inflection at ${\simeq}1.4{\times}10^{19}$\,eV, and the suppression.
This procedure is applied to spectra inferred from a single detector type (i.e.\ water-Cherenkov detectors), but can be used for the combination of any spectral measurements for which the uncorrelated uncertainties can be estimated.

The impressive regularity of the all-particle spectrum observed in the energy region between the second knee and the ankle can hide an underlying intertwining of different astrophysical phenomena, which might be exposed by looking at the spectrum of different primary elements. 
In the future, further measurements will allow separation of the intensities due to the different components.
On the one hand, $X_\text{max}$ values will be determined down to $10^{17}$\,eV using the three HEAT telescopes.
On the other hand, the determination of the muon component of EAS above $10^{17}$\,eV will be possible using the new array of underground muon detectors~\cite{Aab:2020frk}, co-located with the SD-750.
This will help us in studying whether the origin of the second knee stems from, for instance, the steep fall-off of an iron component, as expected for Galactic CRs characterized by a rigidity-dependent maximum acceleration energy for particles with charge $Z$, namely $E_\text{max}(Z)\simeq ZE_\text{max}^\text{proton}$. 
In addition, we will be able to extend the measurement of the energy spectrum below $10^{17}\,$eV with a denser array of 433\,m-spaced detectors and with the analysis of the Cherenkov light in FD events~\cite{Novotny:20194U}.
The extension will allow us to lower the threshold and to further explore the second-knee region in more detail.  

\begin{acknowledgements}

\begin{sloppypar}
The successful installation, commissioning, and operation of the Pierre
Auger Observatory would not have been possible without the strong
commitment and effort from the technical and administrative staff in
Malarg\"ue. We are very grateful to the following agencies and
organizations for financial support:
\end{sloppypar}

\begin{sloppypar}
Argentina -- Comisi\'on Nacional de Energ\'\i{}a At\'omica; Agencia Nacional de
Promoci\'on Cient\'\i{}fica y Tecnol\'ogica (ANPCyT); Consejo Nacional de
Investigaciones Cient\'\i{}ficas y T\'ecnicas (CONICET); Gobierno de la
Provincia de Mendoza; Municipalidad de Malarg\"ue; NDM Holdings and Valle
Las Le\~nas; in gratitude for their continuing cooperation over land
access; Australia -- the Australian Research Council; Belgium -- Fonds
de la Recherche Scientifique (FNRS); Research Foundation Flanders (FWO);
Brazil -- Conselho Nacional de Desenvolvimento Cient\'\i{}fico e Tecnol\'ogico
(CNPq); Financiadora de Estudos e Projetos (FINEP); Funda\c{c}\~ao de Amparo \`a
Pesquisa do Estado de Rio de Janeiro (FAPERJ); S\~ao Paulo Research
Foundation (FAPESP) Grants No.~2019/10151-2, No.~2010/07359-6 and
No.~1999/05404-3; Minist\'erio da Ci\^encia, Tecnologia, Inova\c{c}\~oes e
Comunica\c{c}\~oes (MCTIC); Czech Republic -- Grant No.~MSMT CR LTT18004,
LM2015038, LM2018102, CZ.02.1.01/0.0/0.0/16{\textunderscore}013/0001402,
CZ.02.1.01/0.0/0.0/18{\textunderscore}046/0016010 and
CZ.02.1.01/0.0/0.0/17{\textunderscore}049/0008422; France -- Centre de Calcul
IN2P3/CNRS; Centre National de la Recherche Scientifique (CNRS); Conseil
R\'egional Ile-de-France; D\'epartement Physique Nucl\'eaire et Corpusculaire
(PNC-IN2P3/CNRS); D\'epartement Sciences de l'Univers (SDU-INSU/CNRS);
Institut Lagrange de Paris (ILP) Grant No.~LABEX ANR-10-LABX-63 within
the Investissements d'Avenir Programme Grant No.~ANR-11-IDEX-0004-02;
Germany -- Bundesministerium f\"ur Bildung und Forschung (BMBF); Deutsche
Forschungsgemeinschaft (DFG); Finanzministerium Baden-W\"urttemberg;
Helmholtz Alliance for Astroparticle Physics (HAP);
Helmholtz-Gemeinschaft Deutscher Forschungszentren (HGF); Ministerium
f\"ur Innovation, Wissenschaft und Forschung des Landes
Nordrhein-Westfalen; Ministerium f\"ur Wissenschaft, Forschung und Kunst
des Landes Baden-W\"urttemberg; Italy -- Istituto Nazionale di Fisica
Nucleare (INFN); Istituto Nazionale di Astrofisica (INAF); Ministero
dell'Istruzione, dell'Universit\'a e della Ricerca (MIUR); CETEMPS Center
of Excellence; Ministero degli Affari Esteri (MAE); M\'exico -- Consejo
Nacional de Ciencia y Tecnolog\'\i{}a (CONACYT) No.~167733; Universidad
Nacional Aut\'onoma de M\'exico (UNAM); PAPIIT DGAPA-UNAM; The Netherlands
-- Ministry of Education, Culture and Science; Netherlands Organisation
for Scientific Research (NWO); Dutch national e-infrastructure with the
support of SURF Cooperative; Poland -Ministry of Science and Higher
Education, grant No.~DIR/WK/2018/11; National Science Centre, Grants
No.~2013/08/M/ST9/00322, No.~2016/23/B/ST9/01635 and No.~HARMONIA
5--2013/10/M/ST9/00062, UMO-2016/22/M/ST9/00198; Portugal -- Portuguese
national funds and FEDER funds within Programa Operacional Factores de
Competitividade through Funda\c{c}\~ao para a Ci\^encia e a Tecnologia
(COMPETE); Romania -- Romanian Ministry of Education and Research, the
Program Nucleu within MCI (PN19150201/16N/2019 and PN19060102) and
project PN-III-P1-1.2-PCCDI-2017-0839/19PCCDI/2018 within PNCDI III;
Slovenia -- Slovenian Research Agency, grants P1-0031, P1-0385, I0-0033,
N1-0111; Spain -- Ministerio de Econom\'\i{}a, Industria y Competitividad
(FPA2017-85114-P and PID2019-104676GB-C32), Xunta de Galicia (ED431C
2017/07), Junta de Andaluc\'\i{}a (SOMM17/6104/UGR, P18-FR-4314) Feder Funds,
RENATA Red Nacional Tem\'atica de Astropart\'\i{}culas (FPA2015-68783-REDT) and
Mar\'\i{}a de Maeztu Unit of Excellence (MDM-2016-0692); USA -- Department of
Energy, Contracts No.~DE-AC02-07CH11359, No.~DE-FR02-04ER41300,
No.~DE-FG02-99ER41107 and No.~DE-SC0011689; National Science Foundation,
Grant No.~0450696; The Grainger Foundation; Marie Curie-IRSES/EPLANET;
European Particle Physics Latin American Network; University of Delaware Research Foundation (UDRF) - 2019; and UNESCO.
\end{sloppypar}

\end{acknowledgements}

\appendix
\section{The Electromagnetic Trigger Algorithms}
\label{app::triggeralgo}

\begin{figure}[t]
\centering
\includegraphics[width=0.99\columnwidth]{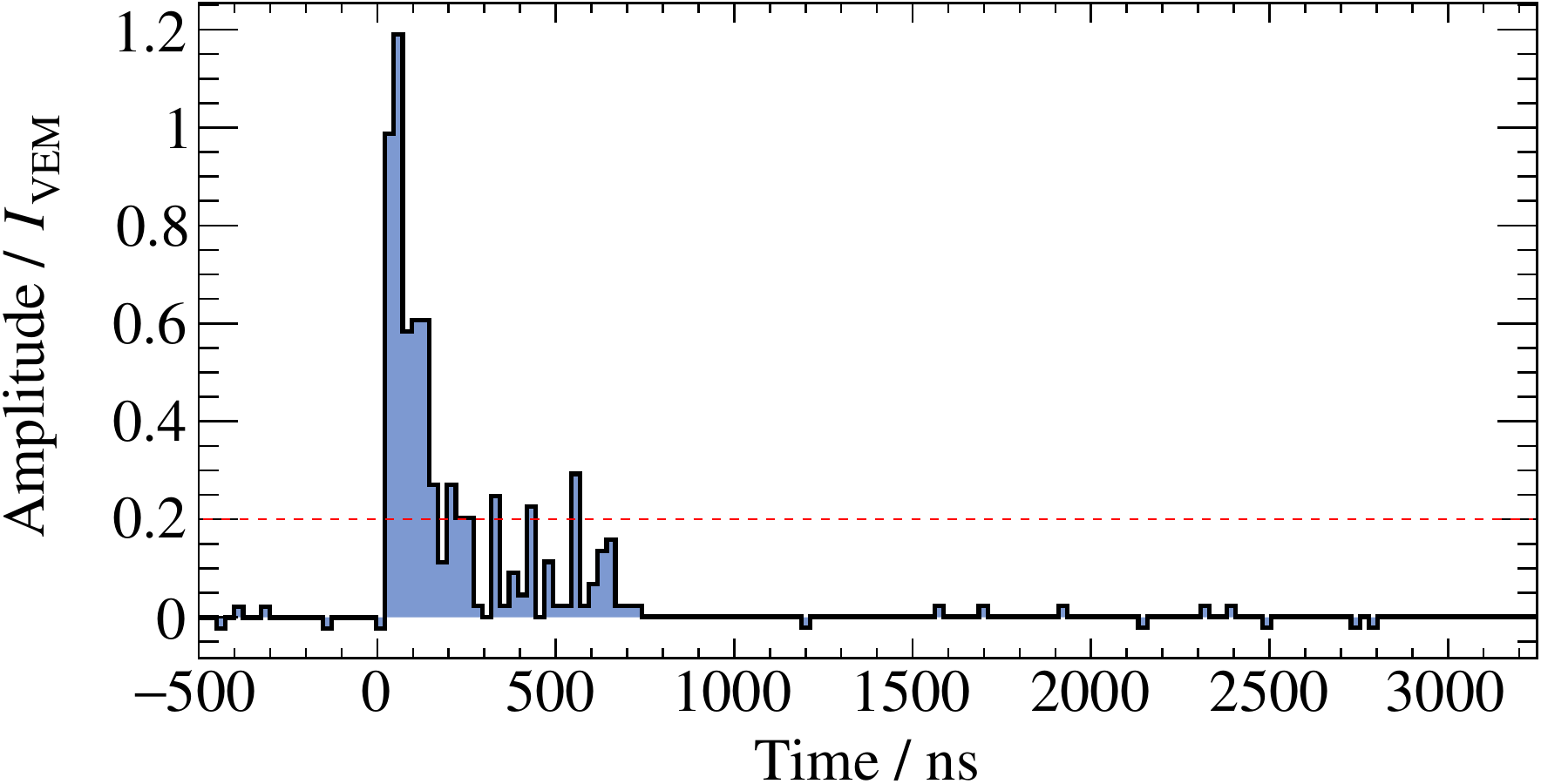}\\
\vspace{0.4cm}
\includegraphics[width=0.99\columnwidth]{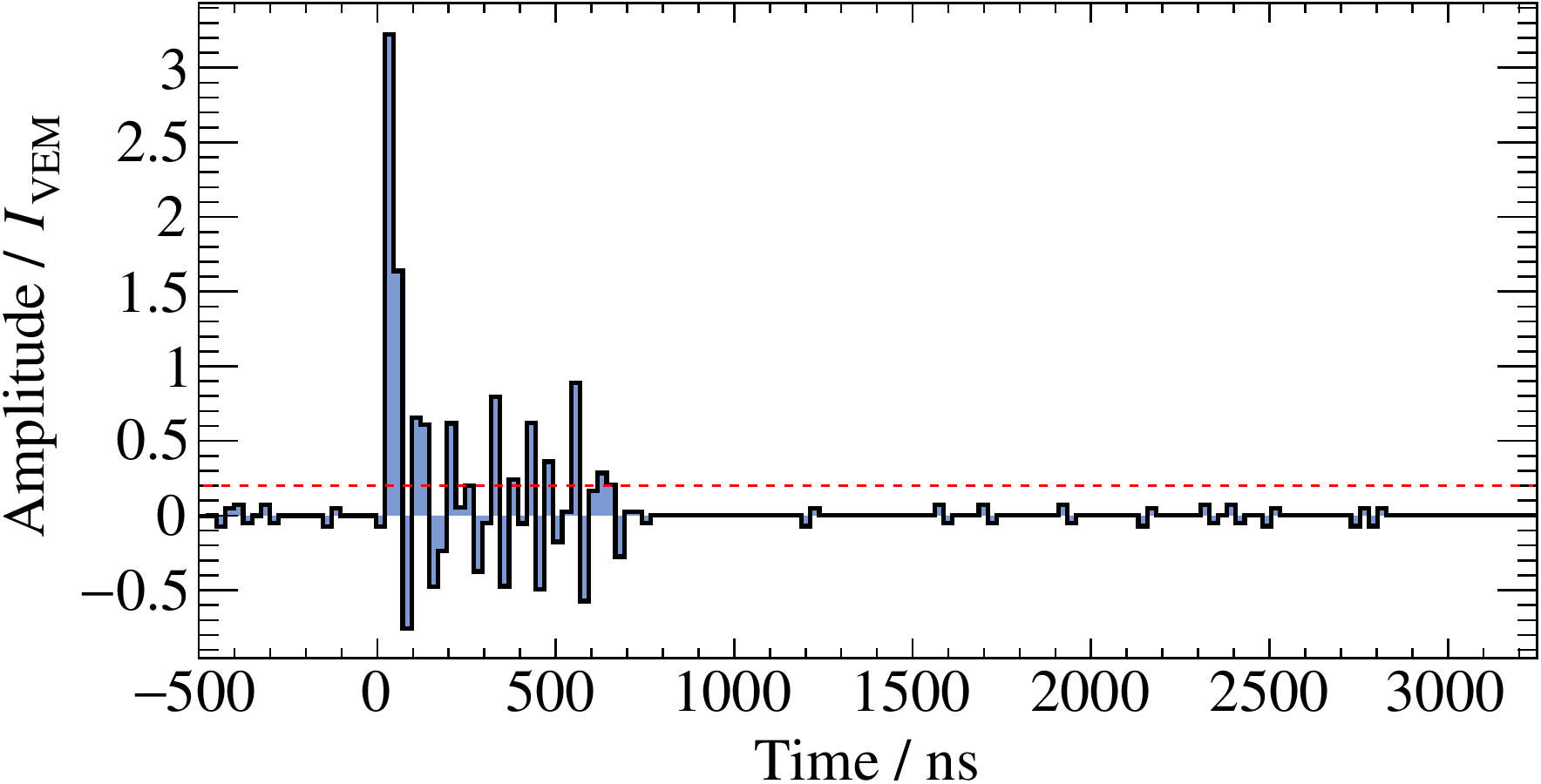}\\
\vspace{0.4cm}
\includegraphics[width=0.99\columnwidth]{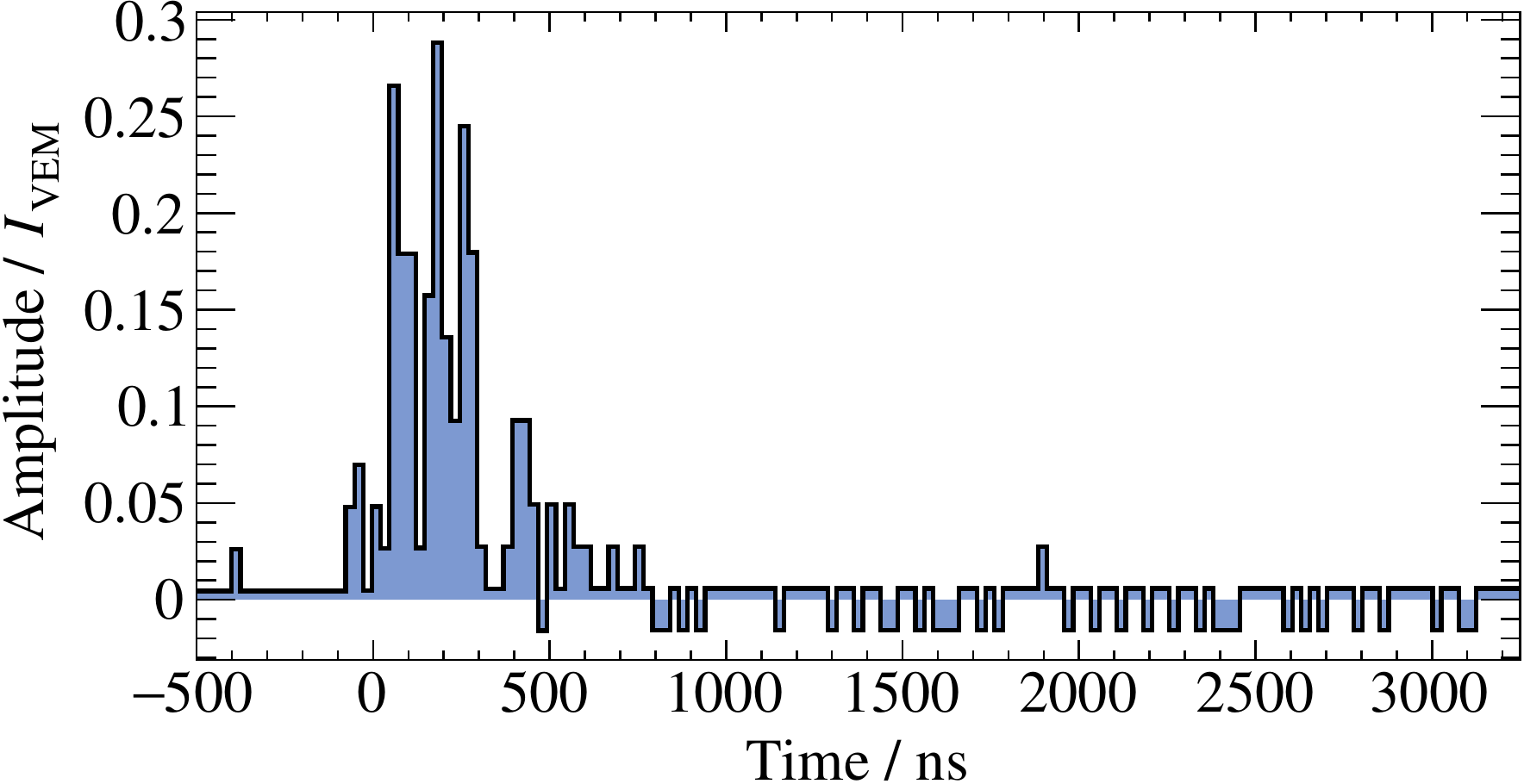}
\caption{Top: Example waveform which passes the ToTd algorithm. Middle: The deconvolution (\cref{eq:TOTdAlgorithm}) of the first waveform along with the threshold to pass the algorithm (dashed red line). Bottom: Example waveform which passes the MoPS algorithm.}
\label{f:ExampleWaveforms}
\end{figure}

The ToTd and MoPS triggers were designed to be insensitive to atmospheric muons such that they enable the detection of small electromagnetic signals from air showers.
The typical morphology of a waveform from a $\sim$GeV muon is a $\simeq$150\,ns ($\simeq$6 ADC bins) pulse with an amplitude of ${\simeq}1\,I_\text{VEM}$, where $I_\text{VEM}$ is the maximum amplitude of a signal created by a muon that traverses the water volume vertically~\cite{Bertou:2005ze}.
Thus, the ToTd and MoPS algorithms are used to look for signals that do not fit this criteria.

The two additional triggers build upon the ToT trigger in two ways, applying more sophisticated analyses to the signal waveform. They are aimed at further suppressing the muon background so as to enhance the sensitivity to pure electromagnetic signals, which are generally smaller.

The ToTd trigger uses the typical decay time of Cherenkov light inside the water volume, $\tau = 67$\,ns, to deconvolve the exponential tail of the pulses before applying the ToT condition. This has the effect of reducing the influence of muons in the trigger, since the typical signal from a muon, with fast rise time and ${\approx}60$\,ns decay constant, is compressed into one or two time bins. The exponential tail of the signal is deconvolved using
\begin{equation}
    D_i = \frac{S_i - S_{i-1} e^{-\Delta t / \tau}}{1 - \mathrm{e}^{-\Delta t / \tau}},
    \label{eq:TOTdAlgorithm}
\end{equation}
where $S_i$ is the signal in the $i$-th time-bin and $\Delta t = 25$\,ns is the ADC bin-width.
For an exponential decay with the mean decay time, the deconvolved values, $D_i$, would be zero. 
However for an exponential decay with statistical noise that is proportional to $\sqrt{S_i}$, the set $\{D_i\}$ would exponentially decrease with an increased decay length $\tau' = 2\tau$.
After performing the deconvolution in \cref{eq:TOTdAlgorithm}, the trigger is satisfied if ${\ge}13$\,ADC bins (${\ge}325$\,ns) are above $0.2\,I_\text{VEM}$, in coincidence between two of the three PMTs, within a sliding 3\,\textmu s (120\,bin) time window. 
An example of a waveform which passes the ToTd trigger and its deconvolution are shown in the top two plots of \cref{f:ExampleWaveforms}.
Only 11 bins are above $0.2\,I_\text{VEM}$ in the original waveform such that it cannot pass the traditional TOT algorithm. However the deconvolution has the 13 bins required to be above the threshold.

The second, MoPS, counts the number of instances, in a sliding 3\,\textmu s window, in which there is a monotonic increase of the signal amplitude. 
Each such instance of successive increases in the digitized waveform is what we define as a \emph{positive step}.\footnote{For example, four bins with $S_i \leq S_{i+1} \leq S_{i+2} \leq S_{i+3}$ is considered one positive step, not three positive steps.}
For each positive step, the total vertical increase, $j$, must be above that of typical noise, and below the characteristic amplitude of a vertical muon, namely $3< j \leq 31$. 
If more than four of the positive-step instances fall within this range, the trigger condition is satisfied. 
An example of a waveform which passes the MoPS trigger is shown in the bottom plot of \cref{f:ExampleWaveforms}.
\section{Spectrum Data}
\label{app::spectrumdata}

We report in this appendix several data of interest. Note that more can be found in the Supplemental Material in electronic format.

\begin{figure}[ht]
\centering
\includegraphics[width=0.99\columnwidth]{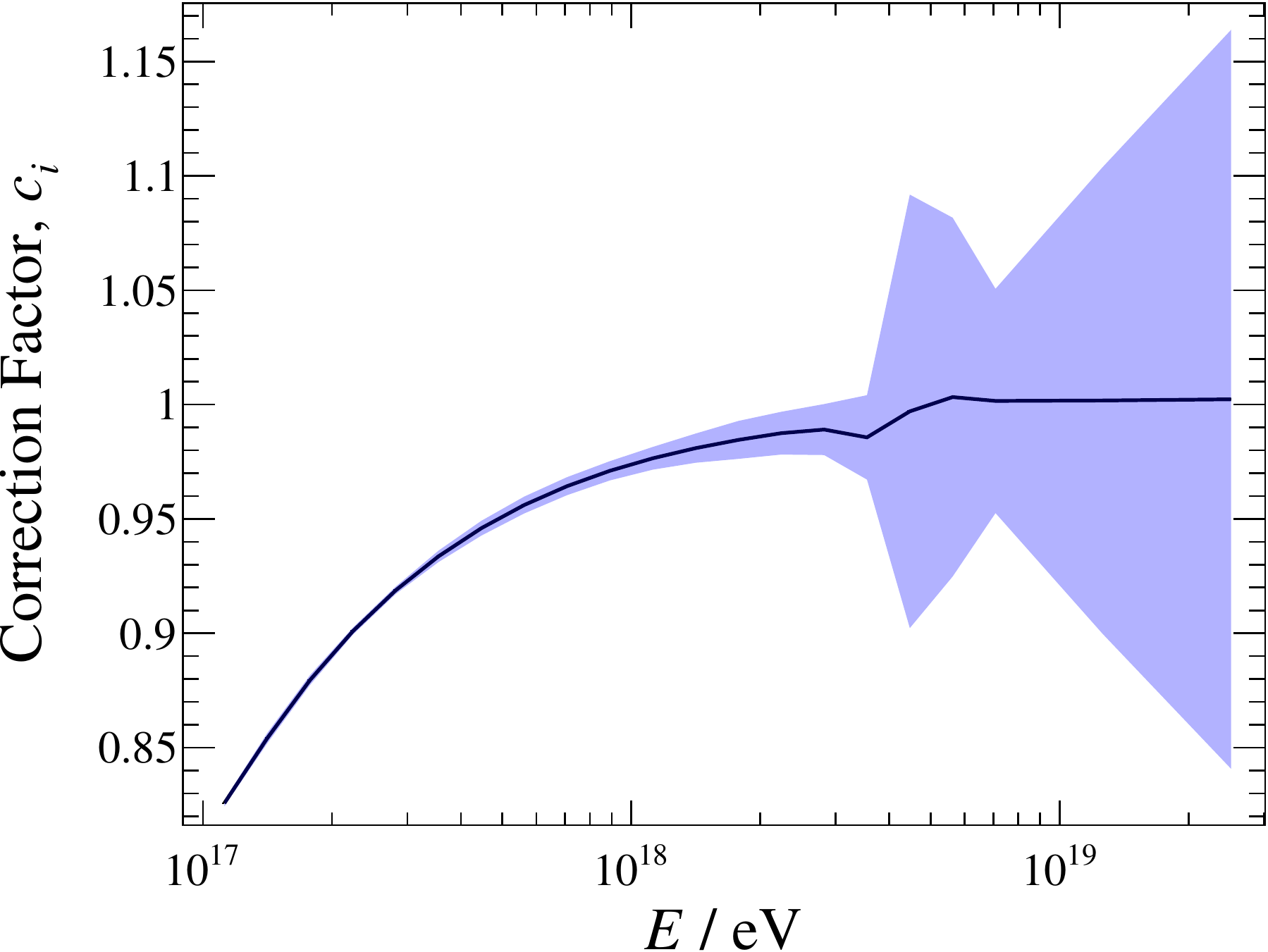}
\caption{The scaling factor that has been applied to the raw spectrum to produce the unfolded spectrum (see \cref{eqn:Junf}) and the statistical uncertainty.}
\label{f:CorrFact}
\end{figure}
The bin migration is corrected to produce the unfolded spectrum. The magnitude of the correction factor, as described by \cref{eqn:Junf}, is shown in \cref{f:CorrFact} along with the statistical uncertainty band.
The energy spectrum of the SD-750 array is reported in \cref{tab:spectrum750} and the correlation matrix of the spectral parameters at the nominal energy scale in \cref{tab:rho_param750stat} (statistical uncertainties).
Finally, the combined energy spectrum is reported in \cref{tab:spectrumSD} and the correlation matrix of the spectral parameters at the nominal energy scale in \cref{tab:rho_paramSDstat} (statistical uncertainties).

\begin{table}[ht]
\caption{\small{SD-750 spectrum data. The correlations between systematic uncertainties are provided in the Supplementary material.}}
\begin{tabular}{l c c}
\hline
  \textbf{$\boldsymbol{\lg(E/\text{eV})}$} & $\boldsymbol N$ & \textbf{$\boldsymbol{\frac{J\pm\sigma_\text{stat}\pm\sigma_\text{syst}}{\text{km}^{2}\,\text{yr}\,\text{sr}\,\text{eV}}}$} \\
\hline
17.05 & 217094 & $\left(6.568~^{+0.015~+2.0}_{-0.015~-1.8}\right){\times} 10^{-14}$ \\
17.15 & 132828 & $\left(3.302~^{+0.010~+1.0}_{-0.010~-0.9}\right){\times} 10^{-14}$ \\
17.25 & 79931 & $\left(1.625~^{+0.006~+0.5}_{-0.006~-0.5}\right){\times} 10^{-14}$ \\
17.35 & 47509 & $\left(7.860~^{+0.038~+2.5}_{-0.038~-2.3}\right){\times} 10^{-15}$ \\
17.45 & 27889 & $\left(3.738~^{+0.023~+1.2}_{-0.023~-1.1}\right){\times} 10^{-15}$ \\
17.55 & 16407 & $\left(1.775~^{+0.014~+0.6}_{-0.014~-0.5}\right){\times} 10^{-15}$ \\
17.65 & 9695 & $\left(8.44~^{+0.09~+2.9}_{-0.09~-2.5}\right){\times} 10^{-16}$ \\ 
17.75 & 5653 & $\left(3.95~^{+0.05~+1.5}_{-0.05~-1.1}\right){\times} 10^{-16}$ \\
17.85 & 3317 & $\left(1.86~^{+0.03~+0.7}_{-0.03~-0.5}\right){\times} 10^{-16}$ \\
17.95 & 1990 & $\left(8.91~^{+0.20~+3.6}_{-0.20~-2.6}\right){\times} 10^{-17}$ \\
18.05 & 1158 & $\left(4.14~^{+0.12~+1.8}_{-0.12~-1.2}\right){\times} 10^{-17}$ \\
18.15 & 651 & $\left(1.85~^{+0.07~+0.9}_{-0.07~-0.5}\right){\times} 10^{-17}$ \\
18.25 & 367 & $\left(8.35~^{+0.43~+4.1}_{-0.45~-2.4}\right){\times} 10^{-18}$ \\
18.35 & 235 & $\left(4.26~^{+0.27~+2.2}_{-0.28~-1.2}\right){\times} 10^{-18}$ \\
18.45 & 139 & $\left(2.01~^{+0.17~+1.1}_{-0.17~-0.6}\right){\times} 10^{-18}$ \\
18.55 & 79 & $\left(9.0~^{+1.0~+5.2}_{-1.0~-2.5}\right){\times} 10^{-19}$ \\
18.65 & 45 & $\left(4.1~^{+0.7~+2.4}_{-0.6~-1.2}\right){\times} 10^{-19}$ \\
18.75 & 31 & $\left(2.3~^{+0.4~+1.3}_{-0.4~-0.6}\right){\times} 10^{-19}$ \\
18.85 & 29 & $\left(1.7~^{+0.3~+0.9}_{-0.3~-0.5}\right){\times} 10^{-19}$ \\
19.10 & 36 & $\left(2.8~^{+0.5~+1.6}_{-0.5~-0.8}\right){\times} 10^{-20}$ \\
19.40 & 7 & $\left(5.7~^{+2.4~+3.2}_{-2.4~-1.6}\right){\times} 10^{-21}$ \\
\hline
\end{tabular}
\label{tab:spectrum750}
\end{table}

\begin{table}[ht]
\caption{\small{Elements of the correlation matrix (statistical uncertainties) of the spectral parameters describing the SD-750 energy spectrum at the nominal energy scale.}}
\begin{tabular}{l c c c c c}
\hline
  & $J_0$ & $\gamma_1$ & $E_{12}$ & $\gamma_2$ & $\omega_{01}$  \\
\hline
$J_0$ & \phantom{-}1\phantom{.000} & $\phantom{-}0.978$ & $-0.067$ & $-0.120$ & $\phantom{-}0.998$ \\
$\gamma_1$ &   & \phantom{-}1\phantom{.000} & $-0.094$ & $-0.109$ & $\phantom{-}0.967$ \\
$E_{12}$ &  &   & \phantom{-}1\phantom{.000} & $-0.814$ & $-0.059$ \\
$\gamma_2$ &  &  &  &  \phantom{-}1\phantom{.000} & $-0.123$ \\
$\omega_{01}$ &  &   &  &   & \phantom{-}1\phantom{.000} \\
\hline
\end{tabular}
\label{tab:rho_param750stat}
\end{table}%

\clearpage

\begin{table}[ht]
\caption{\small{Combined SD spectrum data. The correlations between systematic uncertainties are provided in the Supplementary material.}}
\begin{tabular}{l c}
\hline
\textbf{$\boldsymbol{\lg(E/\text{eV})}$} &  \textbf{$\boldsymbol{\frac{J\pm\sigma_\text{stat}\pm\sigma_\text{syst}}{\text{km}^{2}\,\text{yr}\,\text{sr}\,\text{eV}}}$} \\
\hline
17.05  & $\left(6.341~^{+0.015~+2.1}_{-0.015~-1.9}\right){\times} 10^{-14}$ \\
17.15  & $\left(3.191~^{+0.010~+1.1}_{-0.010~-0.9}\right){\times} 10^{-14}$ \\
17.25  & $\left(1.577~^{+0.006~+0.5}_{-0.006~-0.5}\right){\times} 10^{-14}$ \\
17.35  & $\left(7.643~^{+0.039~+2.6}_{-0.039~-2.3}\right){\times} 10^{-15}$ \\
17.45  & $\left(3.650~^{+0.024~+1.3}_{-0.024~-1.1}\right){\times} 10^{-15}$ \\
17.55  & $\left(1.739~^{+0.015~0.6}_{-0.015~-0.5}\right){\times} 10^{-15}$ \\
17.65  & $\left(8.32~^{+0.09~+3.0}_{-0.09~-2.4}\right){\times} 10^{-16}$ \\ 
17.75  & $\left(3.90~^{+0.05~+1.4}_{-0.05~-1.1}\right){\times} 10^{-16}$ \\
17.85  & $\left(1.85~^{+0.03~+0.7}_{-0.03~-0.5}\right){\times} 10^{-16}$ \\
17.95  & $\left(8.87~^{+0.20~+3.3}_{-0.20~-2.5}\right){\times} 10^{-17}$ \\
18.05  & $\left(4.14~^{+0.12~+1.6}_{-0.12~-1.2}\right){\times} 10^{-17}$ \\
18.15  & $\left(1.90~^{+0.07~+0.7}_{-0.07~-0.5}\right){\times} 10^{-17}$ \\
18.25  & $\left(8.47~^{+0.43~+3.3}_{-0.44~-2.4}\right){\times} 10^{-18}$ \\
18.35  & $\left(4.17~^{+0.28~+1.7}_{-0.27~-1.2}\right){\times} 10^{-18}$ \\
18.45  & $\left(1.929~^{+0.007~+0.7}_{-0.007~-0.5}\right){\times} 10^{-18}$ \\
18.55  & $\left(9.041~^{+0.042~+2.9}_{-0.042~-2.0}\right){\times} 10^{-19}$ \\
18.65  & $\left(4.294~^{+0.026~+1.3}_{-0.026~-0.9}\right){\times} 10^{-19}$ \\
18.75  & $\left(2.167~^{+0.016~+0.6}_{-0.016~-0.4}\right){\times} 10^{-19}$ \\
18.85  & $\left(1.226~^{+0.011~+0.3}_{-0.011~-0.2}\right){\times} 10^{-19}$ \\
18.95  & $\left(6.82~^{+0.08~+1.6}_{-0.08~-1.3}\right){\times} 10^{-20}$ \\
19.05  & $\left(3.79~^{+0.05~+0.9}_{-0.05~-0.7}\right){\times} 10^{-20}$ \\
19.15  & $\left(2.07~^{+0.03~+0.5}_{-0.03~-0.4}\right){\times} 10^{-20}$ \\
19.25  & $\left(1.04~^{+0.02~+0.2}_{-0.02~-0.2}\right){\times} 10^{-20}$ \\
19.35  & $\left(0.53~^{+0.01~+0.16}_{-0.01~-0.13}\right){\times} 10^{-20}$ \\
19.45  & $\left(2.49~^{+0.08~+0.9}_{-0.08~-0.7}\right){\times} 10^{-21}$ \\
19.55  & $\left(1.25~^{+0.05~+0.5}_{-0.05~-0.3}\right){\times} 10^{-21}$ \\
19.65  & $\left(5.99~^{+0.32~+2.4}_{-0.32~-1.8}\right){\times} 10^{-22}$ \\
19.75  & $\left(1.95~^{+0.17~+0.9}_{-0.17~-0.7}\right){\times} 10^{-22}$ \\
19.85  & $\left(8.1~^{+1.0~+4.0}_{-0.9~-2.8}\right){\times} 10^{-23}$ \\
19.95  & $\left(1.8~^{+0.5~+1.0}_{-0.4~-0.7}\right){\times} 10^{-23}$ \\
20.05  & $\left(5.5~^{+2.5~+3.3}_{-1.8~-2.2}\right){\times} 10^{-24}$ \\
20.15  & $\left(2.9~^{+1.7~+1.9}_{-1.2~-1.2}\right){\times} 10^{-24}$ \\
\hline
\end{tabular}
\label{tab:spectrumSD}
\end{table}%

\begin{table*}[ht]
\caption{\small{Elements of the correlation matrix (statistical uncertainties) of the spectral parameters describing the combined SD energy spectrum.}}
\begin{center}
\begin{tabular}{l c c c c c c c c c}
\hline
  & $J_0$ & $\gamma_1$ & $E_{12}$ & $\gamma_2$ & $E_{23}$ & $\gamma_3$ & $E_{34}$ & $\gamma_4$ & $\omega_{01}$  \\
\hline
$J_0$ & \phantom{-}1\phantom{.000} &$-0.470$ & $\phantom{-}0.552$ &$-0.357$ &$-0.383$ &$-0.095$ &$-0.033$ & $\phantom{-}0.035$ &$-0.258$ \\
$\gamma_1$ & & \phantom{-}1\phantom{.000} &$-0.585$ & $\phantom{-}0.524$ & $\phantom{-}0.877$ & $\phantom{-}0.358$ &$\phantom{-}0.075$ &$-0.085$ & $\phantom{-}0.966$ \\
$E_{12}$ &  & & \phantom{-}1\phantom{.000} &$-0.896$ &$-0.425$ &$\phantom{-}0.192$ & $\phantom{-}0.119$ & $\phantom{-}0.110$ &$-0.493$ \\
$\gamma_2$ & & & & \phantom{-}1\phantom{.000} & $\phantom{-}0.455$ & $-0.385$ & $-0.217$ & $-0.154$ & $\phantom{-}0.475$ \\
$E_{23}$ & & & & & \phantom{-}1\phantom{.000} & $\phantom{-}0.252$ & $-0.063$ & $-0.174$ & $\phantom{-}0.858$ \\
$\gamma_3$ & & & & & & \phantom{-}1\phantom{.000} & $\phantom{-}0.474$ & $\phantom{-}0.136$ & $\phantom{-}0.366$ \\
$E_{34}$ & & & & & & & \phantom{-}1\phantom{.000} & $\phantom{-}0.805$ &$\phantom{-}0.075$ \\
$\gamma_4$ & & & & & & & & \phantom{-}1\phantom{.000} &$-0.078$ \\
$\omega_{01}$ & &  & & & & & & & \phantom{-}1\phantom{.000} \\
\hline
\end{tabular}
\label{tab:rho_paramSDstat}
\end{center}
\end{table*}%

\clearpage

\bibliographystyle{JHEP.bst}
\bibliography{main.bib}

\clearpage
\onecolumn
\section*{The Pierre Auger Collaboration}\par

P.~Abreu$^{71}$,
M.~Aglietta$^{53,51}$,
J.M.~Albury$^{12}$,
I.~Allekotte$^{1}$,
A.~Almela$^{8,11}$,
J.~Alvarez-Mu\~niz$^{78}$,
R.~Alves Batista$^{79}$,
G.A.~Anastasi$^{62,51}$,
L.~Anchordoqui$^{86}$,
B.~Andrada$^{8}$
% ...
\\
\noindent\llap{$^{1}$} Centro At\'omico Bariloche and Instituto Balseiro (CNEA-UNCuyo-CONICET), San Carlos de Bariloche, Argentina\\
\llap{$^{2}$} Centro de Investigaciones en L\'aseres y Aplicaciones, CITEDEF and CONICET, Villa Martelli, Argentina\\
\llap{$^{3}$} Departamento de F\'\i{}sica and Departamento de Ciencias de la Atm\'osfera y los Oc\'eanos, FCEyN, Universidad de Buenos Aires and CONICET, Buenos Aires, Argentina\\
\llap{$^{4}$} IFLP, Universidad Nacional de La Plata and CONICET, La Plata, Argentina\\
\llap{$^{5}$} Instituto de Astronom\'\i{}a y F\'\i{}sica del Espacio (IAFE, CONICET-UBA), Buenos Aires, Argentina\\
\llap{$^{6}$} Instituto de F\'\i{}sica de Rosario (IFIR) -- CONICET/U.N.R.\ and Facultad de Ciencias Bioqu\'\i{}micas y Farmac\'euticas U.N.R., Rosario, Argentina\\
\llap{$^{7}$} Instituto de Tecnolog\'\i{}as en Detecci\'on y Astropart\'\i{}culas (CNEA, CONICET, UNSAM), and Universidad Tecnol\'ogica Nacional -- Facultad Regional Mendoza (CONICET/CNEA), Mendoza, Argentina\\
\llap{$^{8}$} Instituto de Tecnolog\'\i{}as en Detecci\'on y Astropart\'\i{}culas (CNEA, CONICET, UNSAM), Buenos Aires, Argentina\\
\llap{$^{9}$} Observatorio Pierre Auger, Malarg\"ue, Argentina\\
\llap{$^{10}$} Observatorio Pierre Auger and Comisi\'on Nacional de Energ\'\i{}a At\'omica, Malarg\"ue, Argentina\\
\llap{$^{11}$} Universidad Tecnol\'ogica Nacional -- Facultad Regional Buenos Aires, Buenos Aires, Argentina\\
\llap{$^{12}$} University of Adelaide, Adelaide, S.A., Australia\\
\llap{$^{13}$} Universit\'e Libre de Bruxelles (ULB), Brussels, Belgium\\
\llap{$^{14}$} Vrije Universiteit Brussels, Brussels, Belgium\\
\llap{$^{15}$} Centro Brasileiro de Pesquisas Fisicas, Rio de Janeiro, RJ, Brazil\\
\llap{$^{16}$} Centro Federal de Educa\c{c}\~ao Tecnol\'ogica Celso Suckow da Fonseca, Nova Friburgo, Brazil\\
\llap{$^{17}$} Instituto Federal de Educa\c{c}\~ao, Ci\^encia e Tecnologia do Rio de Janeiro (IFRJ), Brazil\\
\llap{$^{18}$} Universidade de S\~ao Paulo, Escola de Engenharia de Lorena, Lorena, SP, Brazil\\
\llap{$^{19}$} Universidade de S\~ao Paulo, Instituto de F\'\i{}sica de S\~ao Carlos, S\~ao Carlos, SP, Brazil\\
\llap{$^{20}$} Universidade de S\~ao Paulo, Instituto de F\'\i{}sica, S\~ao Paulo, SP, Brazil\\
\llap{$^{21}$} Universidade Estadual de Campinas, IFGW, Campinas, SP, Brazil\\
\llap{$^{22}$} Universidade Estadual de Feira de Santana, Feira de Santana, Brazil\\
\llap{$^{23}$} Universidade Federal do ABC, Santo Andr\'e, SP, Brazil\\
\llap{$^{24}$} Universidade Federal do Paran\'a, Setor Palotina, Palotina, Brazil\\
\llap{$^{25}$} Universidade Federal do Rio de Janeiro, Instituto de F\'\i{}sica, Rio de Janeiro, RJ, Brazil\\
\llap{$^{26}$} Universidade Federal do Rio de Janeiro (UFRJ), Observat\'orio do Valongo, Rio de Janeiro, RJ, Brazil\\
\llap{$^{27}$} Universidade Federal Fluminense, EEIMVR, Volta Redonda, RJ, Brazil\\
\llap{$^{28}$} Universidad de Medell\'\i{}n, Medell\'\i{}n, Colombia\\
\llap{$^{29}$} Universidad Industrial de Santander, Bucaramanga, Colombia\\
\llap{$^{30}$} Charles University, Faculty of Mathematics and Physics, Institute of Particle and Nuclear Physics, Prague, Czech Republic\\
\llap{$^{31}$} Institute of Physics of the Czech Academy of Sciences, Prague, Czech Republic\\
\llap{$^{32}$} Palacky University, RCPTM, Olomouc, Czech Republic\\
\llap{$^{33}$} CNRS/IN2P3, IJCLab, Universit\'e Paris-Saclay, Orsay, France\\
\llap{$^{34}$} Laboratoire de Physique Nucl\'eaire et de Hautes Energies (LPNHE), Sorbonne Universit\'e, Universit\'e de Paris, CNRS-IN2P3, Paris, France\\
\llap{$^{35}$} Univ.\ Grenoble Alpes, CNRS, Grenoble Institute of Engineering Univ.\ Grenoble Alpes, LPSC-IN2P3, 38000 Grenoble, France\\
\llap{$^{36}$} Universit\'e Paris-Saclay, CNRS/IN2P3, IJCLab, Orsay, France\\
\llap{$^{37}$} Bergische Universit\"at Wuppertal, Department of Physics, Wuppertal, Germany\\
\llap{$^{38}$} Karlsruhe Institute of Technology (KIT), Institute for Experimental Particle Physics, Karlsruhe, Germany\\
\llap{$^{39}$} Karlsruhe Institute of Technology (KIT), Institut f\"ur Prozessdatenverarbeitung und Elektronik, Karlsruhe, Germany\\
\llap{$^{40}$} Karlsruhe Institute of Technology (KIT), Institute for Astroparticle Physics, Karlsruhe, Germany\\
\llap{$^{41}$} RWTH Aachen University, III.\ Physikalisches Institut A, Aachen, Germany\\
\llap{$^{42}$} Universit\"at Hamburg, II.\ Institut f\"ur Theoretische Physik, Hamburg, Germany\\
\llap{$^{43}$} Universit\"at Siegen, Department Physik -- Experimentelle Teilchenphysik, Siegen, Germany\\
\llap{$^{44}$} Gran Sasso Science Institute, L'Aquila, Italy\\
\llap{$^{45}$} INFN Laboratori Nazionali del Gran Sasso, Assergi (L'Aquila), Italy\\
\llap{$^{46}$} INFN, Sezione di Catania, Catania, Italy\\
\llap{$^{47}$} INFN, Sezione di Lecce, Lecce, Italy\\
\llap{$^{48}$} INFN, Sezione di Milano, Milano, Italy\\
\llap{$^{49}$} INFN, Sezione di Napoli, Napoli, Italy\\
\llap{$^{50}$} INFN, Sezione di Roma ``Tor Vergata'', Roma, Italy\\
\llap{$^{51}$} INFN, Sezione di Torino, Torino, Italy\\
\llap{$^{52}$} Istituto di Astrofisica Spaziale e Fisica Cosmica di Palermo (INAF), Palermo, Italy\\
\llap{$^{53}$} Osservatorio Astrofisico di Torino (INAF), Torino, Italy\\
\llap{$^{54}$} Politecnico di Milano, Dipartimento di Scienze e Tecnologie Aerospaziali , Milano, Italy\\
\llap{$^{55}$} Universit\`a del Salento, Dipartimento di Matematica e Fisica ``E.\ De Giorgi'', Lecce, Italy\\
\llap{$^{56}$} Universit\`a dell'Aquila, Dipartimento di Scienze Fisiche e Chimiche, L'Aquila, Italy\\
\llap{$^{57}$} Universit\`a di Catania, Dipartimento di Fisica e Astronomia, Catania, Italy\\
\llap{$^{58}$} Universit\`a di Milano, Dipartimento di Fisica, Milano, Italy\\
\llap{$^{59}$} Universit\`a di Napoli ``Federico II'', Dipartimento di Fisica ``Ettore Pancini'', Napoli, Italy\\
\llap{$^{60}$} Universit\`a di Palermo, Dipartimento di Fisica e Chimica ''E.\ Segr\`e'', Palermo, Italy\\
\llap{$^{61}$} Universit\`a di Roma ``Tor Vergata'', Dipartimento di Fisica, Roma, Italy\\
\llap{$^{62}$} Universit\`a Torino, Dipartimento di Fisica, Torino, Italy\\
\llap{$^{63}$} Benem\'erita Universidad Aut\'onoma de Puebla, Puebla, M\'exico\\
\llap{$^{64}$} Unidad Profesional Interdisciplinaria en Ingenier\'\i{}a y Tecnolog\'\i{}as Avanzadas del Instituto Polit\'ecnico Nacional (UPIITA-IPN), M\'exico, D.F., M\'exico\\
\llap{$^{65}$} Universidad Aut\'onoma de Chiapas, Tuxtla Guti\'errez, Chiapas, M\'exico\\
\llap{$^{66}$} Universidad Michoacana de San Nicol\'as de Hidalgo, Morelia, Michoac\'an, M\'exico\\
\llap{$^{67}$} Universidad Nacional Aut\'onoma de M\'exico, M\'exico, D.F., M\'exico\\
\llap{$^{68}$} Universidad Nacional de San Agustin de Arequipa, Facultad de Ciencias Naturales y Formales, Arequipa, Peru\\
\llap{$^{69}$} Institute of Nuclear Physics PAN, Krakow, Poland\\
\llap{$^{70}$} University of \L{}\'od\'z, Faculty of High-Energy Astrophysics,\L{}\'od\'z, Poland\\
\llap{$^{71}$} Laborat\'orio de Instrumenta\c{c}\~ao e F\'\i{}sica Experimental de Part\'\i{}culas -- LIP and Instituto Superior T\'ecnico -- IST, Universidade de Lisboa -- UL, Lisboa, Portugal\\
\llap{$^{72}$} ``Horia Hulubei'' National Institute for Physics and Nuclear Engineering, Bucharest-Magurele, Romania\\
\llap{$^{73}$} Institute of Space Science, Bucharest-Magurele, Romania\\
\llap{$^{74}$} University Politehnica of Bucharest, Bucharest, Romania\\
\llap{$^{75}$} Center for Astrophysics and Cosmology (CAC), University of Nova Gorica, Nova Gorica, Slovenia\\
\llap{$^{76}$} Experimental Particle Physics Department, J.\ Stefan Institute, Ljubljana, Slovenia\\
\llap{$^{77}$} Universidad de Granada and C.A.F.P.E., Granada, Spain\\
\llap{$^{78}$} Instituto Galego de F\'\i{}sica de Altas Enerx\'\i{}as (IGFAE), Universidade de Santiago de Compostela, Santiago de Compostela, Spain\\
\llap{$^{79}$} IMAPP, Radboud University Nijmegen, Nijmegen, The Netherlands\\
\llap{$^{80}$} Nationaal Instituut voor Kernfysica en Hoge Energie Fysica (NIKHEF), Science Park, Amsterdam, The Netherlands\\
\llap{$^{81}$} Stichting Astronomisch Onderzoek in Nederland (ASTRON), Dwingeloo, The Netherlands\\
\llap{$^{82}$} Universiteit van Amsterdam, Faculty of Science, Amsterdam, The Netherlands\\
\llap{$^{83}$} University of Groningen, Kapteyn Astronomical Institute, Groningen, The Netherlands\\
\llap{$^{84}$} Case Western Reserve University, Cleveland, OH, USA\\
\llap{$^{85}$} Colorado School of Mines, Golden, CO, USA\\
\llap{$^{86}$} Department of Physics and Astronomy, Lehman College, City University of New York, Bronx, NY, USA\\
\llap{$^{87}$} Louisiana State University, Baton Rouge, LA, USA\\
\llap{$^{88}$} Michigan Technological University, Houghton, MI, USA\\
\llap{$^{89}$} New York University, New York, NY, USA\\
\llap{$^{90}$} Pennsylvania State University, University Park, PA, USA\\
\llap{$^{91}$} University of Chicago, Enrico Fermi Institute, Chicago, IL, USA\\
\llap{$^{92}$} University of Delaware, Department of Physics and Astronomy, Bartol Research Institute, Newark, DE, USA\\
\llap{$^{93}$} University of Wisconsin-Madison, Department of Physics and WIPAC, Madison, WI, USA\\
\llap{$^{a}$} Fermi National Accelerator Laboratory, Fermilab, Batavia, IL, USA\\
\llap{$^{b}$} Max-Planck-Institut f\"ur Radioastronomie, Bonn, Germany\\
\llap{$^{c}$} School of Physics and Astronomy, University of Leeds, Leeds, United Kingdom\\
\llap{$^{d}$} Colorado State University, Fort Collins, CO, USA\\
\llap{$^{e}$} now at Hakubi Center for Advanced Research and Graduate School of Science, Kyoto University, Kyoto, Japan\\
\llap{$^{f}$} also at University of Bucharest, Physics Department, Bucharest, Romania

\end{document}